\documentclass[aps,prl,reprint,superscriptaddress, longbibliography]{revtex4-1}
\usepackage{H1}
\usepackage[pdftex,colorlinks=true]{hyperref}
\hypersetup{citecolor = blue}
\usepackage[normalem]{ulem}
\usepackage{color}
\usepackage[usenames,dvipsnames,svgnames,table]{xcolor}
\usepackage{mathtools}
\usepackage{enumerate}

\newcommand{\vb}{v_B}

\newcommand{\lamv}{\lambda(v)}

\begin{document}
\title{Velocity-dependent Lyapunov exponents in many-body quantum,\\  semi-classical, and classical chaos
} 

\author{Vedika Khemani}
\affiliation{\mbox{Department of Physics, Harvard University, Cambridge, MA 02138, USA}}
\author{David A. Huse}
\affiliation{\mbox{Department of Physics, Princeton University, Princeton, NJ 08544, USA}}
\author{Adam Nahum}
\affiliation{\mbox{Theoretical Physics, Oxford University, 1 Keble Road, Oxford OX1 3NP, United Kingdom}}

\begin{abstract}
The exponential growth or decay with time of the out-of-time-order commutator (OTOC) is one widely used diagnostic of many-body chaos in spatially-extended systems.  In studies of many-body classical chaos, it has been noted that one can define a velocity-dependent Lyapunov exponent, $\lambda({\bf v})$, which is the growth or decay rate along rays at that velocity.  We examine the behavior of $\lambda({\bf v})$ 
for a variety of many-body systems, both chaotic and integrable.  The so-called light cone for the spreading of operators is defined by $\lambda({\bf \hat n}v_B({\bf \hat n}))=0$, with a generally direction-dependent butterfly speed $v_B({\bf \hat n})$.  In spatially local systems, $\lamv$ is negative outside the light cone where it takes the form $\lambda(v) \sim -(v-v_B)^{\alpha}$ near $\vb$, with the exponent $\alpha$ taking on various values over the range of systems we examine. The regime inside the light cone with positive Lyapunov exponents may only exist for classical, semi-classical or large-$N$ systems, but not for ``fully quantum'' chaotic systems with strong short-range interactions and local Hilbert space dimensions of order one.
\end{abstract}

\maketitle

One central goal of quantum statistical mechanics is understanding whether and how isolated quantum systems
undergoing reversible unitary time evolution are able to bring themselves to thermal equilibrium \cite{Deutsch,Srednicki, Rigol}.  These questions have recently been brought to the forefront following remarkable developments both in experiments (particularly in cold atomic gases~\cite{BlochRMP, Choi2016,Smith2015,KaufmanEntanglement}) and theory (particularly on the topics of many-body localization~\cite{Anderson58, Basko06,  PalHuse,OganesyanHuse, Znidaric,Imbrie2016}, quantum chaos, and new perspectives from holography~\cite{Maldacena_holo, Witten_holo} relating the  physics of information scrambling in black holes to thermalization in other many-body systems~\cite{HaydenPreskill,SekinoSusskind,HosurYoshida,ShenkerStanfordButterfly,Lashkari,LocalizedShocks,CotlerRM, RobertsStanford,KitaevSYK, SachdevSYK}).  We now understand that while memory of the initial conditions is never lost under unitary dynamics, it can get ``scrambled'' or ``hidden" in experimentally inaccessible nonlocal correlations --- leading to an effective  decoherence that can bring local subsytems to thermal equilibrium.  Indeed, one major upshot of this set of developments has been a crisp formulation of the meaning of ``thermalization" in closed quantum systems in terms of the late time values of local observables. By contrast, an equally precise formulation of many-body ``quantum chaos", and if and how it is distinct from thermalization, is still an active area of research and debate. 

One useful window on many-body quantum chaos is provided by the spreading of initially local operators.  In the Heisenberg picture, an operator $O_0$  initially supported at the origin evolves into an operator ${O_0(t) = U^\dagger(t) O_0 U(t)}$ with appreciable support on a region that grows with time $t$.  The out-of-time-order {\it commutator} (OTOC)~\cite{Lieb72,Larkinotoc,KitaevSYK,chaosbound,HosurYoshida,ShenkerStanfordButterfly,LocalizedShocks,CotlerRM,RobertsStanford,GuQiStanford,GuQi_rcft,StanfordWeakCoupling,PatelDiffusiveMetal,ChowdhuryON,Galitski_lyapunov,DoraMoessner,LuitzScrambling,ProsenWeakChaos,AleinerOTOC,MotrunichTFIM_otoc,FradkinHuse,ChalkerFloquetChaos,FawziScrambling,opspreadAdam, opspreadCurt, TiborCons, KhemaniCons} is a simple measure of the footprint of the spatially spreading operator:
\begin{equation}
{C}({\bf x},t) = \frac{1}{2} \langle [O_0(t), W_{\bf{x}}]^\dagger [O_0(t), W_{\bf{x}}] \rangle~,
\label{eq:otoc}
\end{equation} 
where $W_{\bf x}$ is a local operator at position $\bf x$ and the expectation value is taken in a chosen equilibrium ensemble. If $\bf{x}$ is away from the origin, then at early times $W_{\bf{x}}$ either exactly or approximately commutes with $O_0(t)$ and the OTOC is either zero or exponentially small in $x$.  However, as the operator $O_0(t)$ spreads, ${C}({\bf x},t)$ grows to become of order one inside a``light cone'' that is bounded by a ``front'' that propagates in all directions.  (We assume the two operators are normalized so that the long time average of the OTOC is one.)
We focus here on systems where the front propagates ballistically at a nonzero butterfly speed $v_B$ (touching briefly on strongly disordered systems in which operators spread sub-ballistically \cite{Znidaric, BardarsonPollmannMoore, PVP, VHA, NahumRuhmanHuse}). 
In these systems the OTOC increases rapidly as the front approaches $\bf{x}$, and saturates to a constant after the front has passed. We note that $v_B$ will typically depend on the orientation of the front  \cite{opspreadAdam}, unless spatial symmetries prevent this.  

The OTOC has been studied intensely as an early-to-intermediate time diagnostic of quantum chaos in a variety of quantum models whose time-evolution is generated by either a time-independent Hamiltonian\cite{KitaevSYK, ShenkerStanfordButterfly,LocalizedShocks,CotlerRM,RobertsStanford,GuQiStanford,GuQi_rcft,StanfordWeakCoupling,PatelDiffusiveMetal,ChowdhuryON,AleinerOTOC,
DoraMoessner,Galitski_lyapunov,LuitzScrambling,ProsenWeakChaos,MotrunichTFIM_otoc}, or by Floquet dynamics\cite{FradkinHuse,ChalkerFloquetChaos}, or by a random unitary circuit~\cite{FawziScrambling,HosurYoshida,opspreadAdam, opspreadCurt, TiborCons, KhemaniCons}. This diagnostic is natural since 
a classical limit of the OTOC, with the commutator becoming a Poisson bracket, measures the sensitivity of a classical observable at $({\bf x},t)$ to an infinitesimal perturbation of the initial conditions at the origin.
In classical chaotic systems this sensitivity grows exponentially at late times, at a rate set by a Lyapunov exponent,  $C_{\rm cl}\sim e^{\lambda_L t}$, characteristic of the ``butterfly effect" \footnote{We note that the usual definition of the classical Lyapunov exponent involves averaging the logarithm of the factor by which perturbations grow over initial states and perturbations. This is subtly different from the classical analog of the quantum OTOC where the commutator/Poisson bracket is averaged before taking the logarithm. It is worth exploring in future studies whether or not this difference in definitions has any qualitative consequences\cite{Galitski_lyapunov}.}.

Analogously, an exponential growth with time in ${C}({\bf x},t)$ has been widely used as a metric for \emph{quantum} chaos. Since the OTOC is bounded in the quantum setting (due to unitarity), extracting a quantum Lyapunov exponent requires a small parameter $\epsilon$ such that the OTOC scales as $\epsilon$ at early times: ${C \sim \epsilon\; e^{\lambda_L t}}$. This furnishes a parametrically large time scale ${t_* \sim \frac{1}{\lambda_L}\log(\frac{1}{\epsilon})}$ over which the OTOC can grow exponentially before becoming of order one. In the absence of such a small parameter, the OTOC grows to become of order one in some microscopic order one time scale instead of showing an extended period of exponential growth, leaving no room to define a Lyapunov exponent \cite{chaosbound}. Indeed, all known quantum examples that show an exponential growth at intermediate times work in either a large-$N$ or ``semiclassical" weak scattering limit, where the small parameter $\epsilon$ is either set by $1/N$ or a weak scattering rate\cite{KitaevSYK, ShenkerStanfordButterfly,LocalizedShocks,CotlerRM,RobertsStanford,GuQiStanford,StanfordWeakCoupling,PatelDiffusiveMetal,ChowdhuryON,AleinerOTOC}. Some representative examples of such systems include (i) models that are (or are thought to be) holographically dual to black holes\cite{ShenkerStanfordButterfly,LocalizedShocks,CotlerRM}, including the Sachdev-Ye-Kitaev (SYK) model\cite{SachdevSYK, KitaevSYK} and chains of coupled SYK dots at large $N$\cite{GuQiStanford},
(ii) large-$N$ gauge theories\cite{StanfordWeakCoupling,RobertsStanford,ChowdhuryON} and (iii) weakly interacting disordered systems whose low-temperature dynamics are described, to leading order, via a semiclassical weak scattering theory of quasiparticles\cite{PatelDiffusiveMetal}. Additionally, in spatially local systems, a small parameter $\epsilon$ can be defined for ${\bf x}$ away from the origin, since it takes a time linear in the separation between operators $ t_* \sim|{\bf x}|/\vb$ for a large commutator to build up.

We note that even in systems with a large $t_*$, there can exist parametrically longer time scales for thermalization, whence our characterization of the OTOC as an \emph{intermediate} time diagnostic of quantum chaos that is distinct from probes of thermalization at the longest times. These longer times can include, for example,  a  Thouless time $t_{\rm th} \sim L^2$ probing the slow hydrodynamics of diffusive conserved densities or, alternatively, a time scale set by the the inverse energy level spacing which probes level repulsion (a conventional late-time diagnostic of quantum chaos) and scales exponentially with the volume of the system $t_\delta \sim \exp(V)$.  

\begin{figure}
  \includegraphics[width=\columnwidth]{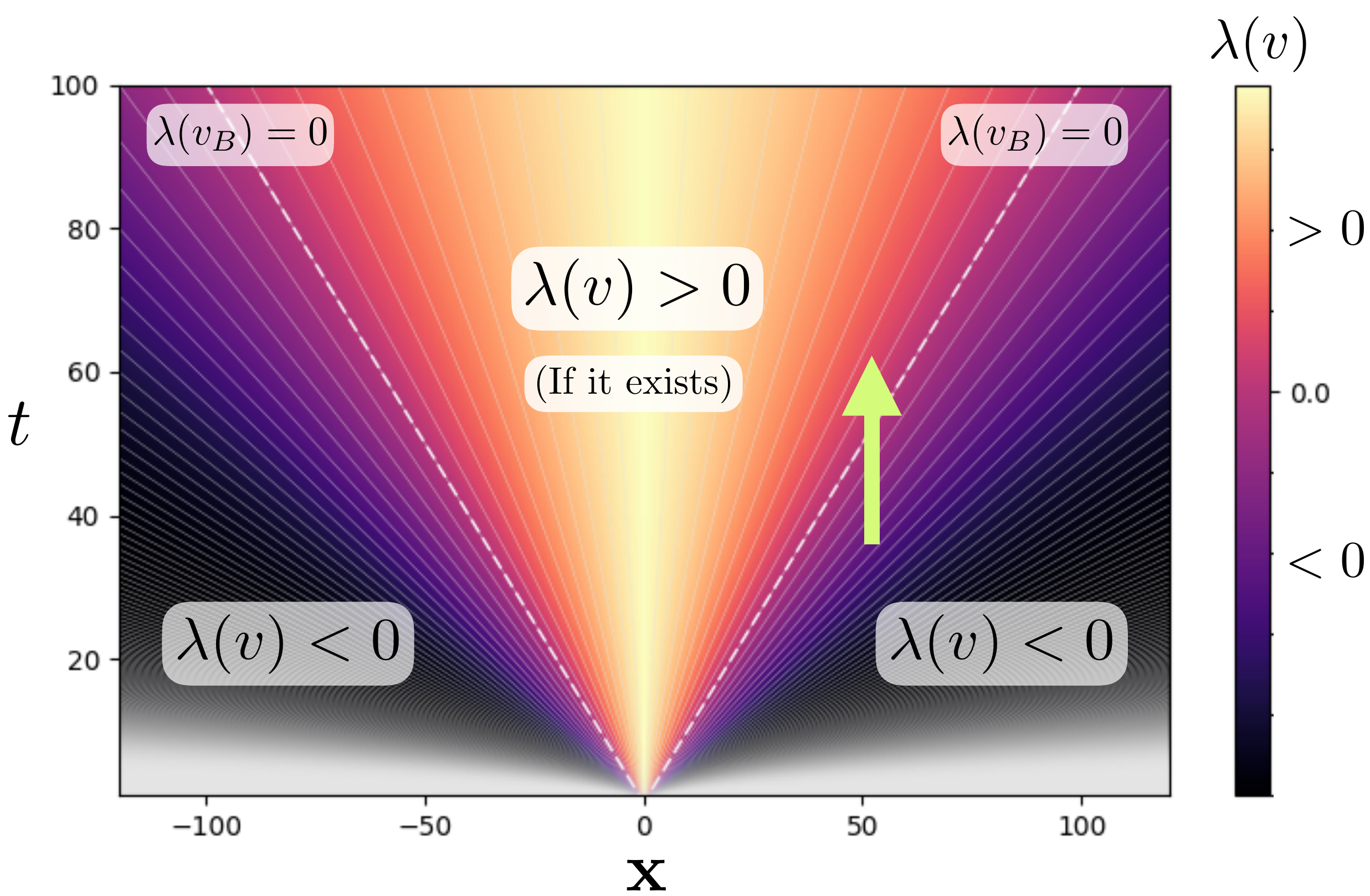}
  \caption{\label{fig:lamv} Schematic illustration of velocity-dependent Lyapunov exponents $\lambda(v)$ along rays $|x|=vt$ in a 1D system. The color scale encodes $\lambda(v)$ which is positive inside the light cone for $v < \vb$ in classical and semiclassical/large-$N$ quantum systems, but is ill-defined inside the cone in ``fully" quantum thermalizing spin systems that do not see an extended period of exponential growth for $v < \vb$. All local systems have  negative $\lambda(v) <0$ for $v > \vb$ following Lieb-Robinson, and $\lambda(v)$ smoothly interpolates from positive to negative in semiclassical/large $N$ systems, passing through zero at $\vb$. The OTOC $C(x_0,t)$ for a fixed position $x_0$ and increasing time (green arrow) cuts through rays with different $\lambda(v)$'s and thus will not show a simple-exponential-in-time growth unless $\lambda(v)$ scales linearly with $v$ \eqref{eq:tdep}.} 
\end{figure}

Here we present a description of the intermediate-time behavior of the OTOC that encompasses many-body quantum, classical and semiclassical chaos within a single framework described by \emph{velocity-dependent Lyapunov exponents} (VDLEs) $\lambda({\bf v})$ \cite{Lieb72,Deissler1984,Kaneko1986,DeisslerKaneko}. These characterize the growth (or decay) with time of the OTOC along rays ${\bf x} = {\bf v}t$, and this framework more completely elucidates the spatiotemporal structure of quantum chaos in spatially extended systems (c.f. Fig.~\ref{fig:lamv}). We define the butterfly velocity $v_B$ in terms of VDLEs and explain the conditions necessary for $C(\mathbf{x}_0,t)$ to display simple-exponential growth with time for fixed $\mathbf{x}_0$.  We find that a careful definition of quantum chaos requires examining the behavior of the OTOC both \emph{inside} and \emph{outside} the ballistic operator front defined by $|{\bf x}| = v_B t$. This allows us to sharply delineate the differences in the behavior of the OTOC between classical/semiclassical/holographic/large-$N$ models and more generic ``fully"  quantum thermalizing spin systems. 

More precisely, only classical and semiclassical/large-$N$ quantum theories have positive VDLEs  for $v< \vb$ \emph{i.e.} exponential growth of the OTOC \emph{inside} the ballistic light cone  $|{\bf x}| < \vb t$ . By contrast,  ``fully'' quantum thermalizing spin systems with strong short-range interactions and small local Hilbert space dimensions (\emph{i.e.} away from semiclassical or large-$N$ limits) lack such a positive exponential growth regime {\it inside} the light cone since such models do not generically have any small parameter $\epsilon$ to prevent the OTOC from saturating on some order one time scale.
On the other hand, if at early times one starts well \emph{outside} the light cone at fixed large ${\bf x_0}$, then spatially local systems do have a large range of growth of the OTOC before the front reaches ${\bf x_0}$. However, even here, the OTOC does \emph{not} show any extended period of simple-exponential-in-time growth in many fully quantum systems (both integrable and non-integrable) since the operator front not only propagates ballistically, but also broadens with  time in such systems~\cite{opspreadAdam, opspreadCurt}.  As a result of this front broadening, fully quantum systems can lack a well-defined positive Lyapunov exponent even outside the ballistic front where a small parameter $\epsilon$ does exist. 

Additionally, all spatially local systems display \emph{negative} VDLEs for  $v > \vb$: these quantify the exponential {decay} of correlations with time along rays {outside} the ballistically propagating front. The existence of negative VDLEs outside the front follows simply from spatial locality and Lieb-Robinson type bounds, and these apply even in nonchaotic integrable systems that display ballistic operator spreading~\cite{CalabreseCardy}.

We posit that the most useful definition(s) of many-body ``quantum chaos" should extend to generic thermalizing spin systems and distinguish between integrable and non-integrable cases. The combination of (i) the lack of a positive $\lambda(v)$ regime in the OTOC for $v < \vb$  in fully quantum chaotic systems and (ii) many qualitative similarities between the early to intermediate time (pre-saturation) growth of the OTOC  between quantum chaotic and integrable systems, suggests that exponential growth in the OTOC may not be best suited as an intermediate time diagnostic of quantum chaos in a fully quantum setting away from semiclassical limits. We note, however, that the saturation behavior of the OTOC for chosen pairs of operators may contain signatures of integrability (on times $t_*\ll t \ll t_\delta$ that are still intermediate). This has so far only been explored in certain cases including free-fermions\cite{MotrunichTFIM_otoc} and conformal field theories\cite{RobertsStanford, GuQi_rcft}, and warrants further study. 
The existence of other quantities~\cite{ProsenWeakChaos} better suited as an intermediate time diagnostic of quantum chaos is an open question (one interesting quantity is the operator entanglement of a spreading operator, which at least in some integrable systems \cite{prosen2007operator,pizorn2009operator,dubail2017entanglement} grows much slower than in chaotic systems \cite{LuitzScrambling, jonay}.)
In all, our results clarify and synthesize several disparate results in the literature, and also provide a useful framework for future studies.  

\emph{Universal Framework ---} We start by describing our general framework and the concept of VDLEs. 
To the best of our knowledge,  VDLEs first make an appearance in the original incarnation of the Lieb-Robinson (LR) bound~\cite{Lieb72}, which stated that for local observables $A, B$ in a local quantum system
\begin{equation}
\lim_{\substack{  {t\rightarrow \infty}\\{ |{\bf x}|>vt}}} ||[A({\bf x},t), B(0,0)]||e^{\mu(v)t} =0,
\label{eq:lr} 
\end{equation}
for all $v> v_{LR}$, with $\mu(v)>0$ a positive increasing function.  The Lieb-Robinson velocity $v_{LR}$ is the minimum speed for which \eqref{eq:lr} holds, and it defines an intrinsic speed of spreading or group-velocity for the system.  The norm of the commutator decays exponentially with time outside the ballistic ``light-cone" $|{\bf x}| = v_{LR}t$, and the function $\mu(v)$ bounds the exponential decay rate along different rays $|{\bf x}| = vt > v_{LR}t $ outside the  cone. 
This velocity-dependent decay rate (which formally is the supremum of all $\mu(v)$ for which \eqref{eq:lr} holds) is the analog of the VDLE as defined below.
This formulation of the LR bound, in terms of $\mu(v)$, emphasizes that the spatiotemporal structure of  operator spreading is organized along rays of fixed velocities in spacetime, a nuance that more recent restatements of the bound~\cite{Nachtergaele1, Nachtergaele2, HastingsSpectralGap} miss by dropping the reference to $\mu(v)$. 

Building on LR and following work on classical chaos~\cite{Deissler1984,Kaneko1986,DeisslerKaneko}, the velocity-dependent Lyapunov exponent $\lambda({\bf v})$ quantifies the exponential growth or decay with time of the OTOC along rays of a given velocity via: 
\begin{equation}
C({\bf x},t) \sim e^{\lambda({\bf v}) t} \;\;  \mbox{ for } \;\; {\bf x} = {\bf v}t~, 
\label{eq:lamvdef}
\end{equation}
and $\lambda({\bf v})$ will in general also depend on the ensemble one is considering (e.g., the temperature for a system with a conserved energy).  
One of the goals of this work is to elucidate the form of $\lambda(v)$ for various physical systems of interest, and to understand when the OTOC exhibits such an exponential growth/decay. 

First, note that the LR bound \eqref{eq:lr} implies that in all local quantum systems, even non-chaotic ones, the OTOC \emph{decays} exponentially with a \emph{negative} VDLE $\lambda(v)$ \emph{outside} the light cone defined by the propagating operator front \emph{i.e.} for $v>\vb$.  
In this language, the butterfly speed $v_B({\bf \hat n})$ along a spatial direction ${\bf \hat n}$ can be defined as the maximum speed $v$ for which 
\begin{equation}
\lambda(v {\bf \hat n})=0~, 
\end{equation}
and this physical propagation speed  is upper-bounded by $v_{LR}$~\cite{RobertsSwingle}.
For simplicity, we will assume full rotational symmetry in the formulas below.
In this case $v_B$ is simply a constant, and is equal to the propagation speed of the front. 
In the absence of rotational symmetry, a subtlety is that  we must distinguish between $\vb({\bf \hat n})$ and the  speed $\widetilde v_B ({\bf \hat n})$ at which a front perpendicular to ${\bf \hat n}$ propagates, but the two speeds are simply related \footnote{Here $\tilde{v}_B({\bf \hat n})$, with a tilde, denotes the  normal propagation speed of a straight front whose normal is parallel to ${\bf \hat n}$. In Ref.~\cite{opspreadAdam} this was denoted $v_B({\bf \hat n})$, but here we use $v_B({\bf \hat n})$ to denote the speed at which  an
initially local operator spreads away from the origin in the direction ${\bf \hat n}$. These differ because  in the absence of rotational symmetry the operator's front is not in general perpendicular to the radial vector, but they are related by a geometrical construction known from classical droplet growth  \cite{WolfWulff, krug1991solids,opspreadAdam}.}.

\begin{figure}[t]
  \includegraphics[width=\columnwidth]{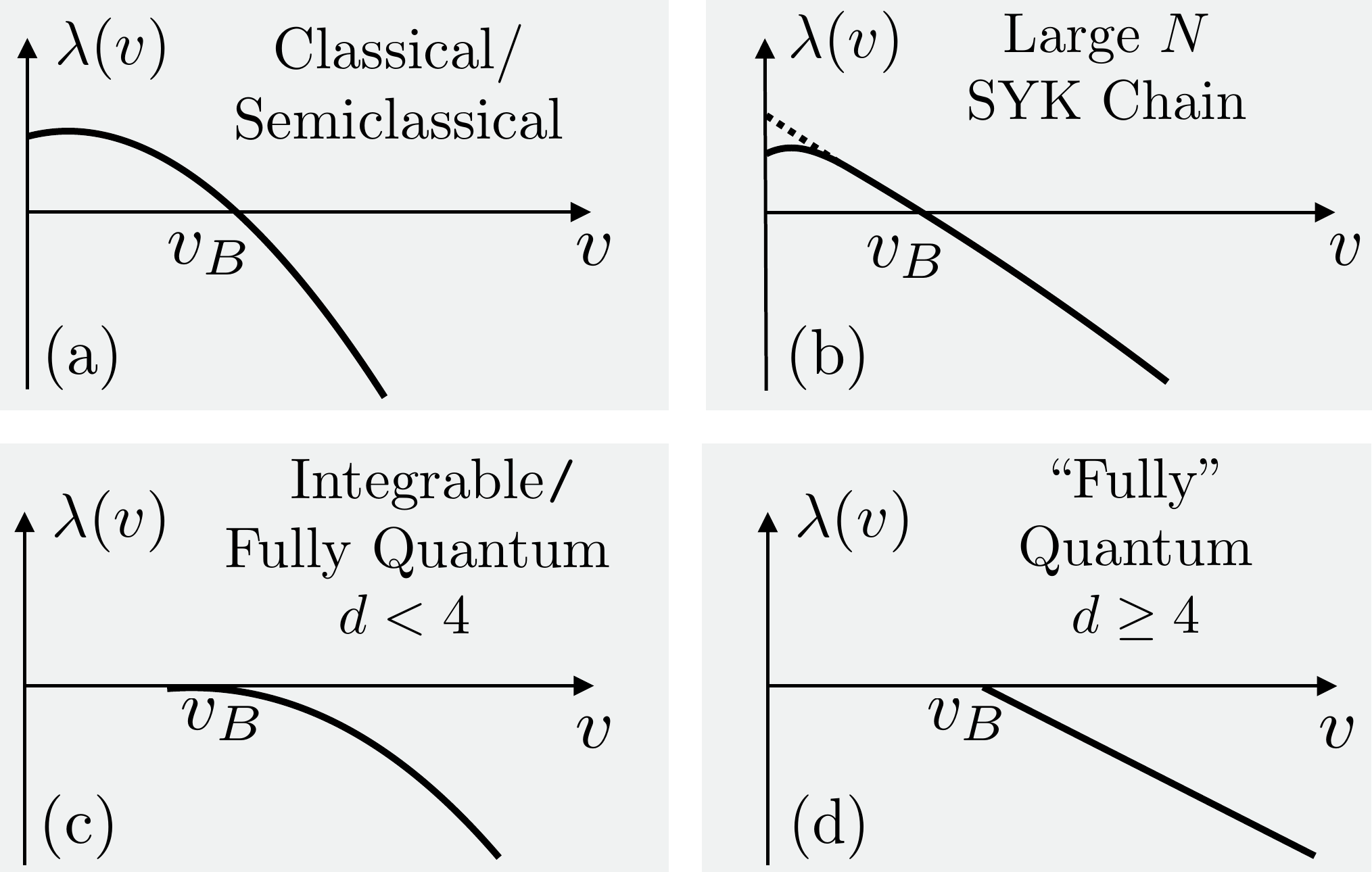}
  \caption{\label{fig:lamvforms} Schematic illustration of $\lambda(v)$ for the different models considered here. (a) In classical and semiclassical weak-scattering systems, $\lamv$ smoothly interpolates from positive to negative, passing through zero at $\vb$. (b) In large $N$ holographic models and chains of coupled SYK dots, the OTOC has a simple exponential form $C \sim \epsilon e^{2\pi T(1-v/\vb)t}$ at low $T$ and leading order in $1/N$. For cases (a) and (b), $\lamv$ scales linearly with $v$ near $\vb$: $\lamv \sim (v-\vb)$, corresponding to $\alpha =1 $ and a simple exponential in time growth for $C(x_0,t)$ near $\vb$. (c) $\lamv$ in fully quantum thermalizing systems with $d<4$ and in integrable models. In these cases, a negative $\lamv$ for $v>\vb$ smoothly approaches zero as $v\rightarrow \vb+$ with exponent $\alpha >1$ due to broadening of the operator front. As a result, these models do not show a simple exponential in time growth in $C(|{\bf x_0}|,t)$ outside the front. (d) In higher dimensions $(d>4)$, the KPZ model capturing the dynamics of operator spreading in chaotic quantum systems has a flat phase which may allow for $\alpha = 1$ and a simple exponential growth in $C(|{\bf x_0}|,t)$ outside the front. Cases (c) and (d) do not show an extended period of exponential growth inside the light cone due to the absence of additional small parameters $\epsilon$, and hence these do not have a well-defined $\lambda(v)$ for $v < \vb$.}
\end{figure}

Next, in chaotic classical and semiclassical/large $N$ many-body systems, the OTOC also \emph{grows} exponentially with time (at least for a parametrically long timescale) along rays  $v< \vb$ as a result of the chaotic growth of local perturbations inside the light cone corresponding to a positive $\lambda(v)$.  This behavior of $\lambda(v)$ in the different regimes is summarized in Fig.~\ref{fig:lamv}.   On the other hand, fully quantum models like spin-1/2 chains do not see an extended period of {exponential} growth inside the light cone --- in such models, the negative $\lambda(v)$ for $v>\vb$ outside of the light cone continuously approaches zero as $v \rightarrow v_B^{+}$ without extending to a positive $\lambda(v)$ regime inside the light cone (Fig.~\ref{fig:lamvforms}(c)).  A similar behavior,  but with some universal differences in the form of $\lambda(v)$ near $v_B$, also occurs in noninteracting and integrable systems, both classical and quantum.

We now comment on the implications of these results for the behavior of $C({\bf x_0},t)$ at fixed ${\bf x_0}$ and increasing $t$ (arrow in  Fig.~\ref{fig:lamv}). For large $x_0$, one is initially outside the light cone and approaches it with increasing $t$, which certainly results in a growth of the OTOC. However, as Fig.~\ref{fig:lamv} shows, this vertical cut passes through rays at different speeds $v$, each with a different decay rate $\lambda(v)$. Thus, in general, 
unless $\lambda(v)$ varies linearly with $v$ for $v \rightarrow v_B^+$ one does not see a simple exponential growth with time in $C({\bf x_0},t)$ just outside the operator front.  
For $v$ larger than and near $\vb$, if we assume that $\lambda(v)$ continuously approaches zero at $\vb$ as a power law with exponent $\alpha$, $\lambda(v) \sim -(v- \vb )^\alpha$, as is the case in all the models we are aware of, then
\begin{align}
C(|{\bf x_0}|=vt, t) &\sim \exp\big({- c (v- \vb )^\alpha}t \big)\nonumber \\
&\sim \exp\left({-c \frac{(|{\bf x_0}|-\vb t)^\alpha}{t^{\alpha-1}}}\right), 
\label{eq:tdep}
\end{align}
which is only a simple exponential in time for $\alpha =1$. 

For chaotic classical/semiclassical/large-$N$ models in which $\lambda(v)$ smoothly interpolates from negative to positive and passes linearly through zero at $\vb$ (Fig.~\ref{fig:lamvforms}(a,b)), the leading term in the Taylor expansion of $\lambda(v)$ near $\vb$ scales linearly with $(v-\vb)$  which is $\alpha=1$, thus giving the known exponential growth with time in $C({\bf x_0},t)$ for these cases for $v \sim \vb$ \emph{both inside and outside the front}. 
However, for fully quantum systems with no additional small parameters $\epsilon$, there is no positive $\lambda(v)$ regime for $v< \vb$ and $\lambda(v)$ can instead can vanish at $v_B$ with exponent $\alpha > 1$ (Fig.~\ref{fig:lamvforms}(c)). This happens whenever the front of the operator broadens with time (see below). In this case there is no simple exponential in time growth of $C({\bf x_0}, t)$ even outside the operator front.  

The relation between $\alpha$ and the width of the (broadening) front follows from an assumption about matching of scaling forms of a standard kind.
Let $\delta x = |{\bf x}| - v_B t$.
If the the front broadens like $t^\beta$ with $\beta>0$, then in the ``typical" regime where $\delta x$ is of order $t^\beta$ we expect that the OTOC may be written as a scaling function of $\delta x / t^\beta$. 
The last line of  \eqref{eq:tdep} resembles such a scaling form 
${f\big( {\delta x}/{t^{\frac{\alpha-1}{\alpha}}}\big)}$
with 
\be
\beta = \frac{\alpha-1}{\alpha}.
\ee
The expression \eqref{eq:tdep} is, however,  valid in a different ``large deviation'' regime in which $\delta x \sim O(t)$ since it applies to $v> \vb$. 
Nevertheless the simplest assumption (which holds in all cases we are aware of) is that as $v \rightarrow \vb^+$ the ``large deviation" scaling function smoothly matches the tail of the ``typical" scaling function for the front region, giving the above relation between exponents. For example a diffusively  broadening front, $\beta = 1/2$, corresponds to $\alpha=2$. We will discuss this matching in more detail below when considering OTOCs in random circuit models.

We now turn to various examples of OTOCs  in classical, quantum and semiclassical/large $N$ systems and examine them through this new lens. 

\emph{Chaotic classical systems---}  The concept of the velocity-dependent (or ``convective" or ``co-moving") Lyapunov exponent was introduced for spatially extended classical chaotic systems in Refs.~\onlinecite{Deissler1984,Kaneko1986,DeisslerKaneko}.  Here we will use a somewhat different definition in order to match to the classical limit of the quantum OTOC:
The average growth or decay rate $\lambda({\bf v})$ of an infinitesimal initial disturbance $\delta u({\bf x}_0=0,t_0=0)$ to a state parameterized by continuous local degrees of freedom $u(x,t)$ is computed in a moving frame with velocity $\bf v$ by calculating 
$|\partial u({\bf x}={\bf v}t,t)/\partial{u(0,0)}|^2$ and averaging over the thermal equilibrium ensemble of initial states: 
\be\label{eq:classicalquantity}
\left\langle \left|\partial u({\bf x}={\bf v}t,t)/\partial{u(0,0)}\right|^2 \right\rangle \sim e^{\lambda({\bf v}) t}.
\ee
In these classical chaotic systems $\lambda({\bf v})$ typically has both a positive and a negative regime, and passes smoothly through zero at a classical butterfly velocity, as illustrated in Fig.~\ref{fig:lamvforms}(a).   This has also recently been explored for a chaotic classical spin chain~\cite{DharClassicalSpinChainChaos} with similar results, and we expect qualitatively similar behavior in higher $d$. 

One way to think about the classical $\lambda({\bf v})$  is by expressing the derivative $\partial u({\bf x}={\bf v}t,t)/\partial{u(0,0)}$ as a sum over paths in spacetime from $(0,0)$ to $({\bf x},t)$ \cite{livi1992scaling,kaneko1992propagation,pikovsky1994roughening}. For illustrative purposes let us discretize time in integer steps: then by the chain rule of differentiation
\be
\f{\partial u({\bf x},t)}{\partial{u(0,0)}}
=
\sum_{{\bf y}_1, \ldots, {\bf y}_{t-1}} W(0,{\bf y}_1, \ldots, {\bf y}_{t-1},{\bf x}),
\ee
where the amplitude $W(0,{\bf y}_1, \ldots, {\bf y}_{t-1},{\bf x})$ for the path
${(0,0)},\, {({\bf y}_1,1)},  \ldots, {({\bf x},t)}$
 is a product of local ``weights'' 
$\partial u({\bf y}_{i+1}, i+1) / \partial u({\bf y}_{i}, i)$ describing how strongly the perturbation is communicated from site ${\bf y}_i$ to site ${\bf y}_{i+1}$ (against the background of a given chaotic configuration).
By locality, this weight will be small if $|{\bf y}_{i+1}- {\bf y}_i|$ is large.
 The exponent $\lambda({\bf v})$
is a measure of how the total weight of paths grows or attenuates with time when their coarse-grained velocity is ${\bf v}$.
It can also be viewed as a ``free energy'' in a statistical ensemble of such paths \footnote{The expression on the left-hand side of (\ref{eq:classicalquantity}) becomes a ``partition function'' for two paths.
 The local weights $\partial u({\bf y}_{i+1}, i+1) / \partial u({\bf y}_{i}, i)$ depend not only on ${\bf y}_{i+1}$ and ${\bf y}_{i}$ but also on the  configuration $u({\bf y_i},i)$. The chaotic time-dependence of $u({\bf y}_i,i)$ means that the configurational average has a similar effect to averaging over weakly correlated randomness in the weights. Since we are averaging the ``partition function'', rather than its logarithm, this is an annealed average, and $-\lambda({\bf v})t$ is an annealed ``free energy'' for the pair of paths. The quenched free energy, in which we take the logarithm before averaging, would give the more conventional definition of the Lyapunov exponent \cite{livi1992scaling,kaneko1992propagation,pikovsky1994roughening}.}. 

\emph{Thermalizing quantum spin models---}  Next, we consider the case of ``fully" quantum thermalizing spin models. 
While exact analytic results for the OTOC are generally not available in these settings for Floquet systems or for time-independent Hamiltonians, recent work has shown that local random unitary circuit models provide a controlled, analytically tractable setting that captures many of the universal aspects of quantum chaos in thermalizing quantum spin systems \cite{nahum2017quantum, opspreadAdam, opspreadCurt, TiborCons, KhemaniCons}.  Refs.~\cite{opspreadAdam, opspreadCurt, TiborCons, KhemaniCons}  have furnished an emergent hydrodynamical picture relating the dynamics of the spreading operator front to stochastic classical surface growth models in one and higher dimensions. From the mapping to noisy surface growth problems, the universal aspects of the front dynamics is captured by the Kardar-Parisi-Zhang (KPZ) equation in $(d-1)$ spatial dimensions~\cite{opspreadAdam}. This gives a front that not only propagates ballistically with speed $\vb$, but also \emph{broadens} in time as $t^\beta$ with $\beta = \frac{1}{2}, \frac{1}{3},  0.241 \pm 0.001$ in $d=1,2,3$, respectively. This leads to a scaling $\lambda(v)\sim -(v-\vb)^{\alpha}$ with $\alpha=\frac{1}{1-\beta} >1$, and hence no simple exponential in time growth in $C({\bf x_0},t)$ \eqref{eq:tdep}.  

Starting with 1d, in a qubit chain model whose dynamics are given by a random unitary local circuit, the (left and right) operator fronts can each be described by a spatial probability distribution for a fictitious biased random walker, and this biased diffusion results in fronts that propagate ballistically with speed $v_B$, but also broaden diffusively with the front width scaling as $\sqrt{t}$ in time~\cite{opspreadAdam, opspreadCurt}.  The behavior of the averaged OTOC at long times (with no additional structure or conservation laws, and reflection symmetry on average) is, in the ``typical'' regime, $|x| - \vb t \sim O(\sqrt{t})$, near the front
\begin{align}
C_{1d}^{\rm rc}(x,t) \sim \f{1}{4} \left[ 1+  {\rm erf} \left( \frac{\vb t +x}{ \sqrt{2Dt} } \right) \right]\left[1+  {\rm erf} \left( \frac{\vb t -x}{ \sqrt{2Dt} } \right) \right],
\label{eq:1drc}
\end{align}
where $D$ is the diffusion constant of the fictitious  walker (not to be confused with the diffusivity of any conserved observables that are transported within the system) and the superscript ${\rm rc}$ refers to ``random circuit". 
Outside the front in the large deviation regime, but for $v$ close to $v_B$, the exact microscopic probability \cite{opspreadAdam, opspreadCurt} for the random walker to travel an improbably large distance $vt$ leads to
\begin{align}
C_{1d}^{\rm rc}(x,t) 
\sim \exp\left[{-\frac{(|x|-\vb t)^2}{2Dt}}\right] = \exp \left[{-\frac{(v-\vb )^2}{2D} t}\right],
\label{eq:1drc_out} 
\end{align}
corresponding to a negative VDLE:
\be
\lambda(v) \approx -\f{(v-\vb)^2}{2D}.
\label{eq:1d_lambda}
\ee
Note that the large deviation form \eqref{eq:1drc_out}  matches smoothly onto the tail of  typical OTOC \eqref{eq:1drc} in the regime $v_B t \gg (|x|-v_B t) \gg \sqrt{2Dt}$, consistent with our discussion below \eqref{eq:tdep}. This qualitative form for $\lambda(v)$ is depicted in Fig.~\ref{fig:lamvforms}(c), with no positive $\lambda(v)$ regime inside the front. 
\footnote{{Inside the light cone there is a large deviation form governing convergence to the saturation value:   
${C_{1d}^{\rm rc}(x,t) \sim  1- \exp\big({-\frac{(v-\vb )^2}{2D} t}\big)}$. 
The exponent here is the continuation of $\lambda(v)$ outside the front. However, in the higher dimensional examples, the large deviation form inside the front scales with a distinct power of $t$, $t^d$ in $d$ spatial dimensions \cite{MajumdarKPZTail}.
In the presence of additional conserved densities (like energy or charge), the late time saturation of the OTOC is a power-law in time instead of exponential~\cite{KhemaniCons, TiborCons}.}}.

As an aside, note that if we view the mapping of the operator dynamics to the 1d diffusion process in terms of the worldline of the random walker,  $-\lambda(v)$ can be viewed as the ``free energy'' cost per unit time for this worldline ``string'' when it is stretched to have speed $v$. This is reminiscent of the relation between $\lambda(v)$ in classical systems and paths in spacetime which we reviewed above, although the two statistical ensembles of paths are quite different. (The random circuit calculations also suggest that the negative $\lambda(v)$ outside the front is related to the ``entanglement line tension'' \cite{jonay} for the second Renyi entropy \footnote{{
In random circuits related random walk pictures underlie the calculation of both the OTOC and the second Renyi entropy \cite{opspreadAdam, opspreadCurt}. 
In these random systems this yields a  relation between $\lambda(v)$ and the ``entanglement line tension'' defined in \cite{jonay}, specifically the line tension $\mathcal{E}_2(v)$ for the second Renyi entropy. This motivates the conjecture, for non-random systems, that $\lambda(v)|_\text{cont} = - s_\text{eq} ( \mathcal{E}_2(v) - v)$, where $s_\text{eq}$ is the thermal entropy density. The left hand side denotes the analytic continuation of $\lambda(v)$ from ${v>v_B}$ to values $v<v_B$. In random circuits  we must distinguish different kinds of averages. The line tension extracted from a calculation of $\overline{e^{-S_2}}$ determines $\lambda(v)$ for  the average OTOC $\overline{C(x,t)}$ by the above formula. It is natural to expect that the line tension determined by the more natural direct average  $\overline{S_2}$  determines $\lambda(v)$ for the typical value of the OTOC,  $\exp \overline{\ln C(x,t)}$. The average and typical values of the OTOC are parametrically close in the region close to the front, but they may differ significantly in the far-front regime where both are exponentially small.}}.)

While thermalizing Hamiltonian spin chains do not show simple exponential in time \emph{growth} for $C({\bf x_0},t)$, we have numerically checked that a negative $\lambda(v)$ \emph{decay} regime consistent with \eqref{eq:1d_lambda} can be found for such systems, as discussed above.
Hamiltonian spin chains also allow $\lambda(v)$ to be defined for arbitrarily large $v$ 
 (in contrast to local quantum circuits, and relativistic field theories, where there is a strict ``light cone'' beyond which even exponentially weak signalling is impossible). A naive perturbative argument suggests that as $v\rightarrow \infty$ the VDLE will scale as 
\begin{align}
\lambda(v) &\sim - v \ln v ,
&
v&\rightarrow \infty.
\end{align}

Turning to 2d, the  behavior of the OTOC near the front follows from a mapping to a cluster growth problem: $C({\bf x},t)$ is proportional to the probability that site ${\bf x}$ has been incorporated inside a growing cluster in a fictitious stochastic growth process \cite{opspreadAdam}. If the effects of lattice anisotropies can be neglected (i.e. in the presence of  rotational invariance
\footnote{{In some circuit models in $d>1$ (which do not have continuous spatial rotation symmetry) some sections of the operator's front can be ``glued'' to the strict lightcone defined by the discrete time circuit \cite{opspreadAdam}. This is a peculiar case where ${v_B( {\bf \hat n} ) = v_\text{LC}( {\bf \hat n})}$ for some directions ${\bf \hat n}$ in space, so that no nontrivial  $\lambda({\mathbf v})$ can be defined for these directions of ${\bf v}$.}}) the Tracy-Widom distribution $F_2$ is expected for the ``typical" scaling:
\begin{align}
C_{2d}^{\rm rc}(r,t) \approx 1- F_2 \left( \frac{r-\vb t}{ ct^{1/3 }} \right) ,
\label{eq:2drc}
\end{align}
where $r$ is a radial coordinate in 2d and $c$ is a constant. 
Large deviations  of the KPZ growth process \cite{MajumdarKPZTail, MonthusKPZNumerics,KolokolovKPZtails} have been studied extensively.  A fluctuation in which a point on the boundary of the cluster travels beyond the mean position by $(v-v_B)t$ requires $O(t)$ rare events, so that $C_{2d}^{\rm rc}(r,t)\sim e^{-t|\lambda(v)|}$ as expected.  The large deviation form has been calculated explicitly in various classical growth models and shown to match smoothly onto the tail of the typical Tracy Widom distribution \eqref{eq:2drc} \cite{MajumdarKPZTail}:
\be\label{eq:2plus1lambda}
\lambda(v) = -\f{4(v-\vb)^{3/2}}{3c^{3/2}}.
\ee
The 3d case is similar but with a different exponent $\beta = 0.241 \pm 0.001$ \cite{KPZ_2d_exponent, KellingKPZ_2d} and scaling form $F\left(\frac{r-\vb t}{t^\beta}\right)$ \cite{HealyScaling1, HealyScaling2}.
It has been numerically verified  \cite{MonthusKPZNumerics,KPZcocktail}, that the tail of this distribution  for large argument decays with exponent $\frac{1}{(1-\beta)}$, consistent with the expected matching.

The  width $t^\beta$ of the operator front  decreases with increasing spatial dimension  corresponding to reduced quantum
fluctuations in the operator front for higher dimensions,
and the exponent $\alpha = \frac{1}{1-\beta}$ decreases towards one, bringing the  form of $C({\bf x_0},t)$ outside the front closer to a simple exponential in time \eqref{eq:tdep}.  In fact, for $d>3$ the  KPZ equation  has a flat phase, suggesting that fronts with $\alpha = 1$ may be possible without fine-tuning for fully quantum systems in $d>3$, see Fig.~\ref{fig:lamvforms}(d). A simple exponential growth with time would then occur in the OTOC outside the front.
Again, this would not continue into a positive $\lambda(v)$ regime inside the front due to the lack of a small parameter $\epsilon$ when $v < \vb$ (Fig.~\ref{fig:lamvforms}(d)). In an appropriate regime, the ``traveling combustion wave'' picture of Ref.~\cite{AleinerOTOC}
also yields  a form with $\lambda(v) = 0$ inside the lightcone and linear behaviour $\alpha=1$ outside the lightcone.

\emph{Integrable models---}We now turn to non-chaotic integrable systems where our presentation will be quite brief. These have no positive $\lamv$ regime, neither in the classical setting nor in the quantum one. Consider first a translationally invariant free fermion lattice system in 1d with quasiparticles of momentum ${k}$ and dispersion $\epsilon_k$. The OTOC between two local fermion densities is readily understood by considering the spreading of a local wave-packet in first quantized notation: $\psi(x = vt ,t) = \int \frac{dk}{2\pi} e^{i (k v - \epsilon_k) t}$. The butterfly velocity $\vb$ is the maximum group-velocity $\vb = \max_k (\frac{\partial \epsilon_k}{\partial k})$ and this defines a $k_B$.  Considering $v \approx \vb$ at late times $t$ and doing a saddle point approximation gives $\psi(x  ,t) \sim \frac{1}{t^{1/3}}{\rm Ai}\left(\frac{x-\vb t}{c t^{1/3}}\right)$ where $\rm{Ai}$ is the Airy function and this form follows because $\epsilon_k$ has an inflection point at $\vb$: $\frac{\partial^2 \epsilon_k}{\partial k^2}|_{k_B} = 0$, and hence evaluating the integral requires going to cubic order in the expansion of $\epsilon_k$ near $k_B$. Thus, for this non-interacting fermion system,
\be
\lambda(v) \sim -(v-\vb)^{3/2}
\ee
for $v>\vb$ which smoothly approaches zero as $v\rightarrow \vb^+$ with $\alpha = 3/2$. The same argument (with the same exponent) holds in any dimension $d$. Likewise, a calculation of $\lambda(v)$ outside the light cone in a non-interacting transverse field Ising model again gives ${\alpha = 3/2}$ for OTOCs involving both the transverse and the longitudinal spin operators~\cite{JacobLesik_private}. This is noteworthy because, although the Ising model maps to free fermions, the longitudinal spin carries a non-local Jordan-Wigner string in the fermion language. 
Thus, growth of the OTOC outside the light cone in free fermion systems is qualitatively similar to the case of non-integrable spin chains, although the two can be 
distinguished  by the value of  $\alpha$ --- except in 2d where by \eqref{eq:2plus1lambda} the exponents happen to be the same (coincidentally, since the mechanisms for front propagation are different). We note, however, that the late-time behavior of the OTOC (inside the light cone) is operator dependent and can decay to zero in non-interacting models~\cite{MotrunichTFIM_otoc} signaling a lack of chaos. 


On the other hand, recent work on \emph{interacting} integrable models in 1d has shown that these have operator fronts that propagate ballistically but spread \emph{diffusively} so that $\alpha = 2$,  coinciding with the value in chaotic 1d systems~\cite{ghkv}. In interacting integrable systems, operators spread through the ballistic propagation of quasiparticles, as in free systems, but with generalized dispersion and momentum relations. The operator front moves with a velocity that is locally set by the fastest quasiparticle velocity as in the non-interacting case. However, in interacting integrable systems, this velocity also depends on the density of the other quasiparticles, so equilibrium density fluctuations cause the front to follow a biased random walk, and therefore to broaden diffusively. Thus, although the mechanisms for operator spreading are distinct in interacting integrable and non-integrable cases, these coarse grained measures of the operator front do not distinguish between the two.

\emph{Semiclassical/ large $N$ models---}  We  now  turn to cases where a large-$N$/semiclassical weak-scattering limit is used to produce a regime of
exponential growth (positive $\lambda(v)$) of the OTOC inside the light cone, with  $\alpha = 1$ near $v_B$ (Fig.~\ref{fig:lamvforms}(a,b)).

The OTOC can be explicitly calculated in various holographic\cite{KitaevSYK,ShenkerStanfordButterfly,LocalizedShocks,CotlerRM,GuQiStanford} and field theoretic\cite{RobertsStanford,ChowdhuryON} models at large $N$ (which includes the case of a chain of coupled SYK dots\cite{GuQiStanford}) and in weak coupling models\cite{StanfordWeakCoupling,PatelDiffusiveMetal}. In several of these models, the OTOC  takes a simple exponential form ${C(x,t) \sim \epsilon e^{\lambda_0 (t- |{\bf x}|/\vb)}}$ with ${\lambda(v) = \lambda_0(1-v/v_B)}$ scaling linearly with $v$. In particular, in the SYK chain model\cite{GuQiStanford}, it is found that over a range of $v$ that includes $v_B$, $\lambda_0$ is at the chaos bound of $2\pi T$.  At low $|v| \ll v_B$ there is also a small regime where $\lambda(v)$ is slightly below this bound, as illustrated in Fig. \ref{fig:lamvforms}(b)\cite{GuQiStanford}.  Another common scaling form for the OTOC, particularly in weakly scattering, weakly disordered models looks like $C(x,t) \sim e^{\lambda_0 t} e^{-{\bf x}^2/(4Dt)}$\cite{PatelDiffusiveMetal}. This corresponds to $\lambda(v) = \lambda_0 -v^2/(4D)$ which defines $\vb = \sqrt{4D\lambda_0}$ (Fig~\ref{fig:lamvforms}(b)). $\lambda_0$ must again obey the chaos bound, and both $\lambda_0$ and $\vb$ tend to zero with decreasing temperature.

  The cases described above with a positive $\lambda(v)$ that is linear near $v_B$ are presumably only valid strictly in the limit of large $N$ or the semiclassical limit, and once quantum effects are fully included in models that are not strictly in such a limit, the long-time behavior of $\lambda(v)$ just outside of the light cone but close to $v_B$ will be as we discuss above for fully quantum models, assuming that the scaling forms that have been conjectured on the basis of random circuits are indeed generic.  For finite $N$, the positive $\lambda(v)$ is then ``only'' an intermediate time behavior, thus making $\lambda(v)$ less sharply defined in this regime. As $N$ is reduced to of order one, this intermediate regime may no longer have a well-defined positive $\lambda(v)$ within the light cone.

\emph{Disordered thermalizing systems---} To end, we briefly consider disordered systems where operator spreading can be sub-ballistic. In disordered 1d chains that are on the thermal side of the many body localization transition (but close to this transition) operator spreading is hindered by weak links with a power-law distribution of timescales \cite{PVP, VHA, NahumRuhmanHuse}. Varying the strength of disorder yields various regimes, including a regime with finite $v_B$ where ${\lambda(v)\sim -(v-v_B)^\alpha}$ and  $\alpha\geq 2$ varies continuously with disorder, and a strong disorder regime with $v_B=0$ and $\lambda\sim - |v|^\alpha$ \footnote{Let the probability distribution for weak-link ``waiting times'' be ${P(\tau)\sim \tau^{-a-2}}$.
At weak disorder $(1<a)$ the broadening of the operator's front \cite{NahumRuhmanHuse} is diffusive, as in the clean system. At intermediate disorder (${0<a<1}$) the front broadens more strongly, giving ${\lambda(v)\sim -(v-v_B)^{(a+1)/a}}$.
For strong disorder ($-1<a<0$) the butterfly speed vanishes: in this regime ${\lambda(v)\sim - |v|^{1-|a|}}$. In the disordered system the definition of $\lambda(v)$ depends on whether we consider e.g. the mean or the typical value of the OTOC, but this should not change these exponents.}.

\emph{Summary---} Many-body quantum chaos is a rich subject in its infancy where many ideas for characterizing notions of ``chaos" and ``scrambling" are still being developed. In this work, we introduce the framework of velocity dependent Lyapunov exponents $\lambda(v)$ for many-body quantum chaos, and use it to more completely characterize the intermediate time behaviour of the OTOC in a variety of different models. 
We find that the spatiotemporal structure of the butterfly effect is organized along rays $|{\bf x}|= vt$ with fixed speeds $v$, with local systems having an intrinsic (possibly direction dependent) spreading speed $\vb$ defined by $\lambda(\vb)=0$. In the quantum setting, only models with a small parameter inhibiting OTOC growth for a long time (for example $1/N$ or a weak scattering rate) 
 display an extended period of exponential growth \emph{inside} the lightcone, i.e. a time regime during which we may define positive VDLEs $\lamv$ for $v <\vb$.  
``Fully quantum'' models with small local Hilbert space dimensions and local interactions do not display an extended period of exponential growth for $v < \vb$. In such models, the OTOC saturates to an order one value in some order one time after the front reaches a particular position. 

On the other hand, \emph{outside} the light-cone, locality  guarantees a negative $\lambda(v)$ regardless of the chaos properties of a system, taking the form ${\lamv \sim -(v-\vb)^\alpha}$ just outside the light-cone. This gives exponential ``Lyapunov--like'' growth only if $\alpha=1$. This is prevented by front-broadening for low-dimensional spin systems, which leads to $\alpha>1$. More studies of the growth and saturation of the OTOC in interacting integrable systems, as well as chaotic ones, to confirm the the similarities and differences in exponents and scaling forms is an important direction for future work.  In all, our work delineates some of the challenges associated with using the intermediate time behavior of the OTOC as a general diagnostic for many-body quantum chaos. Whether one can find other more useful intermediate-time metrics for  chaos, that both extend to ``fully" quantum systems and are distinct from late-time diagnostics of thermalization, remains an open challenge.

\vspace{2pt}
\emph{Related Work:} While we were completing this manuscript, a study~\cite{Swingle_otocMPS} conjecturing the form of the OTOC in Eq.~\eqref{eq:tdep} appeared on the arXiv. Instead, our work derives this form within a broader universal framework of velocity-dependent Lyapunov exponents.

\section*{Acknowledgements}
We thank Xiao Chen, Abhishek Dhar,  Jeongwan Haah, Timothy Halpin-Healy, 
Cheryne Jonay, 
Cheng Ju Lin, Satya Majumdar, Lesik Motrunich, Xiaoliang Qi, Jonathan Ruhman,  Douglas Stanford, Sagar Vijay, and Tianci Zhou for helpful discussions and for previous collaborations.  VK was supported by the Harvard Society of Fellows and the William F. Milton Fund.  
AN acknowledges EPSRC Grant No.~EP/N028678/1. 

\bibliography{global}

\begin{thebibliography}{88}%
\makeatletter
\providecommand \@ifxundefined [1]{%
 \@ifx{#1\undefined}
}%
\providecommand \@ifnum [1]{%
 \ifnum #1\expandafter \@firstoftwo
 \else \expandafter \@secondoftwo
 \fi
}%
\providecommand \@ifx [1]{%
 \ifx #1\expandafter \@firstoftwo
 \else \expandafter \@secondoftwo
 \fi
}%
\providecommand \natexlab [1]{#1}%
\providecommand \enquote  [1]{``#1''}%
\providecommand \bibnamefont  [1]{#1}%
\providecommand \bibfnamefont [1]{#1}%
\providecommand \citenamefont [1]{#1}%
\providecommand \href@noop [0]{\@secondoftwo}%
\providecommand \href [0]{\begingroup \@sanitize@url \@href}%
\providecommand \@href[1]{\@@startlink{#1}\@@href}%
\providecommand \@@href[1]{\endgroup#1\@@endlink}%
\providecommand \@sanitize@url [0]{\catcode `\\12\catcode `\$12\catcode
  `\&12\catcode `\#12\catcode `\^12\catcode `\_12\catcode `\%12\relax}%
\providecommand \@@startlink[1]{}%
\providecommand \@@endlink[0]{}%
\providecommand \url  [0]{\begingroup\@sanitize@url \@url }%
\providecommand \@url [1]{\endgroup\@href {#1}{\urlprefix }}%
\providecommand \urlprefix  [0]{URL }%
\providecommand \Eprint [0]{\href }%
\providecommand \doibase [0]{http://dx.doi.org/}%
\providecommand \selectlanguage [0]{\@gobble}%
\providecommand \bibinfo  [0]{\@secondoftwo}%
\providecommand \bibfield  [0]{\@secondoftwo}%
\providecommand \translation [1]{[#1]}%
\providecommand \BibitemOpen [0]{}%
\providecommand \bibitemStop [0]{}%
\providecommand \bibitemNoStop [0]{.\EOS\space}%
\providecommand \EOS [0]{\spacefactor3000\relax}%
\providecommand \BibitemShut  [1]{\csname bibitem#1\endcsname}%
\let\auto@bib@innerbib\@empty
\bibitem [{\citenamefont {Deutsch}(1991)}]{Deutsch}%
  \BibitemOpen
  \bibfield  {author} {\bibinfo {author} {\bibfnamefont {J.~M.}\ \bibnamefont
  {Deutsch}},\ }\bibfield  {title} {\enquote {\bibinfo {title} {Quantum
  statistical mechanics in a closed system},}\ }\href {\doibase
  10.1103/PhysRevA.43.2046} {\bibfield  {journal} {\bibinfo  {journal} {Phys.
  Rev. A}\ }\textbf {\bibinfo {volume} {43}},\ \bibinfo {pages} {2046--2049}
  (\bibinfo {year} {1991})}\BibitemShut {NoStop}%
\bibitem [{\citenamefont {Srednicki}(1994)}]{Srednicki}%
  \BibitemOpen
  \bibfield  {author} {\bibinfo {author} {\bibfnamefont {Mark}\ \bibnamefont
  {Srednicki}},\ }\bibfield  {title} {\enquote {\bibinfo {title} {Chaos and
  quantum thermalization},}\ }\href {\doibase 10.1103/PhysRevE.50.888}
  {\bibfield  {journal} {\bibinfo  {journal} {Phys. Rev. E}\ }\textbf {\bibinfo
  {volume} {50}},\ \bibinfo {pages} {888--901} (\bibinfo {year}
  {1994})}\BibitemShut {NoStop}%
\bibitem [{\citenamefont {Rigol}\ \emph {et~al.}(2008)\citenamefont {Rigol},
  \citenamefont {Dunjko},\ and\ \citenamefont {Olshanii}}]{Rigol}%
  \BibitemOpen
  \bibfield  {author} {\bibinfo {author} {\bibfnamefont {Marcos}\ \bibnamefont
  {Rigol}}, \bibinfo {author} {\bibfnamefont {Vanja}\ \bibnamefont {Dunjko}}, \
  and\ \bibinfo {author} {\bibfnamefont {Maxim}\ \bibnamefont {Olshanii}},\
  }\bibfield  {title} {\enquote {\bibinfo {title} {{Thermalization and its
  mechanism for generic isolated quantum systems}},}\ }\href {\doibase
  10.1038/nature06838} {\bibfield  {journal} {\bibinfo  {journal} {Nature}\
  }\textbf {\bibinfo {volume} {452}},\ \bibinfo {pages} {854--858} (\bibinfo
  {year} {2008})}\BibitemShut {NoStop}%
\bibitem [{\citenamefont {Bloch}\ \emph {et~al.}(2008)\citenamefont {Bloch},
  \citenamefont {Dalibard},\ and\ \citenamefont {Zwerger}}]{BlochRMP}%
  \BibitemOpen
  \bibfield  {author} {\bibinfo {author} {\bibfnamefont {Immanuel}\
  \bibnamefont {Bloch}}, \bibinfo {author} {\bibfnamefont {Jean}\ \bibnamefont
  {Dalibard}}, \ and\ \bibinfo {author} {\bibfnamefont {Wilhelm}\ \bibnamefont
  {Zwerger}},\ }\bibfield  {title} {\enquote {\bibinfo {title} {Many-body
  physics with ultracold gases},}\ }\href {\doibase 10.1103/RevModPhys.80.885}
  {\bibfield  {journal} {\bibinfo  {journal} {Rev. Mod. Phys.}\ }\textbf
  {\bibinfo {volume} {80}},\ \bibinfo {pages} {885--964} (\bibinfo {year}
  {2008})}\BibitemShut {NoStop}%
\bibitem [{\citenamefont {Choi}\ \emph {et~al.}(2016)\citenamefont {Choi},
  \citenamefont {Hild}, \citenamefont {Zeiher}, \citenamefont {Schau{\ss}},
  \citenamefont {Rubio-Abadal}, \citenamefont {Yefsah}, \citenamefont
  {Khemani}, \citenamefont {Huse}, \citenamefont {Bloch},\ and\ \citenamefont
  {Gross}}]{Choi2016}%
  \BibitemOpen
  \bibfield  {author} {\bibinfo {author} {\bibfnamefont {Jae-yoon}\
  \bibnamefont {Choi}}, \bibinfo {author} {\bibfnamefont {Sebastian}\
  \bibnamefont {Hild}}, \bibinfo {author} {\bibfnamefont {Johannes}\
  \bibnamefont {Zeiher}}, \bibinfo {author} {\bibfnamefont {Peter}\
  \bibnamefont {Schau{\ss}}}, \bibinfo {author} {\bibfnamefont {Antonio}\
  \bibnamefont {Rubio-Abadal}}, \bibinfo {author} {\bibfnamefont {Tarik}\
  \bibnamefont {Yefsah}}, \bibinfo {author} {\bibfnamefont {Vedika}\
  \bibnamefont {Khemani}}, \bibinfo {author} {\bibfnamefont {David~A.}\
  \bibnamefont {Huse}}, \bibinfo {author} {\bibfnamefont {Immanuel}\
  \bibnamefont {Bloch}}, \ and\ \bibinfo {author} {\bibfnamefont {Christian}\
  \bibnamefont {Gross}},\ }\bibfield  {title} {\enquote {\bibinfo {title}
  {Exploring the many-body localization transition in two dimensions},}\ }\href
  {\doibase 10.1126/science.aaf8834} {\bibfield  {journal} {\bibinfo  {journal}
  {Science}\ }\textbf {\bibinfo {volume} {352}},\ \bibinfo {pages} {1547--1552}
  (\bibinfo {year} {2016})}\BibitemShut {NoStop}%
\bibitem [{\citenamefont {{Smith}}\ \emph {et~al.}(2015)\citenamefont
  {{Smith}}, \citenamefont {{Lee}}, \citenamefont {{Richerme}}, \citenamefont
  {{Neyenhuis}}, \citenamefont {{Hess}}, \citenamefont {{Hauke}}, \citenamefont
  {{Heyl}}, \citenamefont {{Huse}},\ and\ \citenamefont
  {{Monroe}}}]{Smith2015}%
  \BibitemOpen
  \bibfield  {author} {\bibinfo {author} {\bibfnamefont {J.}~\bibnamefont
  {{Smith}}}, \bibinfo {author} {\bibfnamefont {A.}~\bibnamefont {{Lee}}},
  \bibinfo {author} {\bibfnamefont {P.}~\bibnamefont {{Richerme}}}, \bibinfo
  {author} {\bibfnamefont {B.}~\bibnamefont {{Neyenhuis}}}, \bibinfo {author}
  {\bibfnamefont {P.~W.}\ \bibnamefont {{Hess}}}, \bibinfo {author}
  {\bibfnamefont {P.}~\bibnamefont {{Hauke}}}, \bibinfo {author} {\bibfnamefont
  {M.}~\bibnamefont {{Heyl}}}, \bibinfo {author} {\bibfnamefont {D.~A.}\
  \bibnamefont {{Huse}}}, \ and\ \bibinfo {author} {\bibfnamefont
  {C.}~\bibnamefont {{Monroe}}},\ }\bibfield  {title} {\enquote {\bibinfo
  {title} {{Many-body localization in a quantum simulator with programmable
  random disorder}},}\ }\href@noop {} {\bibfield  {journal} {\bibinfo
  {journal} {ArXiv e-prints}\ } (\bibinfo {year} {2015})},\ \Eprint
  {http://arxiv.org/abs/1508.07026} {arXiv:1508.07026 [quant-ph]} \BibitemShut
  {NoStop}%
\bibitem [{\citenamefont {{Kaufman}}\ \emph {et~al.}(2016)\citenamefont
  {{Kaufman}}, \citenamefont {{Tai}}, \citenamefont {{Lukin}}, \citenamefont
  {{Rispoli}}, \citenamefont {{Schittko}}, \citenamefont {{Preiss}},\ and\
  \citenamefont {{Greiner}}}]{KaufmanEntanglement}%
  \BibitemOpen
  \bibfield  {author} {\bibinfo {author} {\bibfnamefont {A.~M.}\ \bibnamefont
  {{Kaufman}}}, \bibinfo {author} {\bibfnamefont {M.~E.}\ \bibnamefont
  {{Tai}}}, \bibinfo {author} {\bibfnamefont {A.}~\bibnamefont {{Lukin}}},
  \bibinfo {author} {\bibfnamefont {M.}~\bibnamefont {{Rispoli}}}, \bibinfo
  {author} {\bibfnamefont {R.}~\bibnamefont {{Schittko}}}, \bibinfo {author}
  {\bibfnamefont {P.~M.}\ \bibnamefont {{Preiss}}}, \ and\ \bibinfo {author}
  {\bibfnamefont {M.}~\bibnamefont {{Greiner}}},\ }\bibfield  {title} {\enquote
  {\bibinfo {title} {{Quantum thermalization through entanglement in an
  isolated many-body system}},}\ }\href {\doibase 10.1126/science.aaf6725}
  {\bibfield  {journal} {\bibinfo  {journal} {Science}\ }\textbf {\bibinfo
  {volume} {353}},\ \bibinfo {pages} {794--800} (\bibinfo {year} {2016})},\
  \Eprint {http://arxiv.org/abs/1603.04409} {arXiv:1603.04409 [quant-ph]}
  \BibitemShut {NoStop}%
\bibitem [{\citenamefont {Anderson}(1958)}]{Anderson58}%
  \BibitemOpen
  \bibfield  {author} {\bibinfo {author} {\bibfnamefont {P.~W.}\ \bibnamefont
  {Anderson}},\ }\bibfield  {title} {\enquote {\bibinfo {title} {Absence of
  diffusion in certain random lattices},}\ }\href {\doibase
  10.1103/PhysRev.109.1492} {\bibfield  {journal} {\bibinfo  {journal} {Phys.
  Rev.}\ }\textbf {\bibinfo {volume} {109}},\ \bibinfo {pages} {1492--1505}
  (\bibinfo {year} {1958})}\BibitemShut {NoStop}%
\bibitem [{\citenamefont {Basko}\ \emph {et~al.}(2006)\citenamefont {Basko},
  \citenamefont {Aleiner},\ and\ \citenamefont {Altshuler}}]{Basko06}%
  \BibitemOpen
  \bibfield  {author} {\bibinfo {author} {\bibfnamefont {D.~M.}\ \bibnamefont
  {Basko}}, \bibinfo {author} {\bibfnamefont {I.~L.}\ \bibnamefont {Aleiner}},
  \ and\ \bibinfo {author} {\bibfnamefont {B.~L.}\ \bibnamefont {Altshuler}},\
  }\bibfield  {title} {\enquote {\bibinfo {title} {Metal-insulator transition
  in a weakly interacting many-electron system with localized single-particle
  states},}\ }\href {\doibase 10.1016/j.aop.2005.11.014} {\bibfield  {journal}
  {\bibinfo  {journal} {Annals of Physics}\ }\textbf {\bibinfo {volume}
  {321}},\ \bibinfo {pages} {1126--1205} (\bibinfo {year} {2006})}\BibitemShut
  {NoStop}%
\bibitem [{\citenamefont {Pal}\ and\ \citenamefont {Huse}(2010)}]{PalHuse}%
  \BibitemOpen
  \bibfield  {author} {\bibinfo {author} {\bibfnamefont {Arijeet}\ \bibnamefont
  {Pal}}\ and\ \bibinfo {author} {\bibfnamefont {David~A.}\ \bibnamefont
  {Huse}},\ }\bibfield  {title} {\enquote {\bibinfo {title} {Many-body
  localization phase transition},}\ }\href {\doibase
  10.1103/PhysRevB.82.174411} {\bibfield  {journal} {\bibinfo  {journal} {Phys.
  Rev. B}\ }\textbf {\bibinfo {volume} {82}},\ \bibinfo {pages} {174411}
  (\bibinfo {year} {2010})}\BibitemShut {NoStop}%
\bibitem [{\citenamefont {Oganesyan}\ and\ \citenamefont
  {Huse}(2007)}]{OganesyanHuse}%
  \BibitemOpen
  \bibfield  {author} {\bibinfo {author} {\bibfnamefont {Vadim}\ \bibnamefont
  {Oganesyan}}\ and\ \bibinfo {author} {\bibfnamefont {David~A.}\ \bibnamefont
  {Huse}},\ }\bibfield  {title} {\enquote {\bibinfo {title} {Localization of
  interacting fermions at high temperature},}\ }\href {\doibase
  10.1103/PhysRevB.75.155111} {\bibfield  {journal} {\bibinfo  {journal} {Phys.
  Rev. B}\ }\textbf {\bibinfo {volume} {75}},\ \bibinfo {pages} {155111}
  (\bibinfo {year} {2007})}\BibitemShut {NoStop}%
\bibitem [{\citenamefont {{{\v Z}nidari{\v c}}}\ \emph
  {et~al.}(2008)\citenamefont {{{\v Z}nidari{\v c}}}, \citenamefont
  {{Prosen}},\ and\ \citenamefont {{Prelov{\v s}ek}}}]{Znidaric}%
  \BibitemOpen
  \bibfield  {author} {\bibinfo {author} {\bibfnamefont {M.}~\bibnamefont {{{\v
  Z}nidari{\v c}}}}, \bibinfo {author} {\bibfnamefont {T.}~\bibnamefont
  {{Prosen}}}, \ and\ \bibinfo {author} {\bibfnamefont {P.}~\bibnamefont
  {{Prelov{\v s}ek}}},\ }\bibfield  {title} {\enquote {\bibinfo {title}
  {{Many-body localization in the Heisenberg XXZ magnet in a random field}},}\
  }\href {\doibase 10.1103/PhysRevB.77.064426} {\bibfield  {journal} {\bibinfo
  {journal} {\prb}\ }\textbf {\bibinfo {volume} {77}},\ \bibinfo {eid} {064426}
  (\bibinfo {year} {2008})},\ \Eprint {http://arxiv.org/abs/0706.2539}
  {arXiv:0706.2539 [quant-ph]} \BibitemShut {NoStop}%
\bibitem [{\citenamefont {Imbrie}(2016)}]{Imbrie2016}%
  \BibitemOpen
  \bibfield  {author} {\bibinfo {author} {\bibfnamefont {John~Z.}\ \bibnamefont
  {Imbrie}},\ }\bibfield  {title} {\enquote {\bibinfo {title} {On many-body
  localization for quantum spin chains},}\ }\href {\doibase
  10.1007/s10955-016-1508-x} {\bibfield  {journal} {\bibinfo  {journal}
  {Journal of Statistical Physics}\ }\textbf {\bibinfo {volume} {163}},\
  \bibinfo {pages} {998--1048} (\bibinfo {year} {2016})}\BibitemShut {NoStop}%
\bibitem [{\citenamefont {{Maldacena}}(1999)}]{Maldacena_holo}%
  \BibitemOpen
  \bibfield  {author} {\bibinfo {author} {\bibfnamefont {J.}~\bibnamefont
  {{Maldacena}}},\ }\bibfield  {title} {\enquote {\bibinfo {title} {{The
  Large-N Limit of Superconformal Field Theories and Supergravity}},}\ }\href
  {\doibase 10.1023/A:1026654312961} {\bibfield  {journal} {\bibinfo  {journal}
  {International Journal of Theoretical Physics}\ }\textbf {\bibinfo {volume}
  {38}},\ \bibinfo {pages} {1113--1133} (\bibinfo {year} {1999})}\BibitemShut
  {NoStop}%
\bibitem [{\citenamefont {{Witten}}(1998)}]{Witten_holo}%
  \BibitemOpen
  \bibfield  {author} {\bibinfo {author} {\bibfnamefont {E.}~\bibnamefont
  {{Witten}}},\ }\bibfield  {title} {\enquote {\bibinfo {title} {{Anti-de
  Sitter space and holography}},}\ }\href@noop {} {\bibfield  {journal}
  {\bibinfo  {journal} {Advances in Theoretical and Mathematical Physics}\
  }\textbf {\bibinfo {volume} {2}},\ \bibinfo {pages} {253--291} (\bibinfo
  {year} {1998})},\ \Eprint {http://arxiv.org/abs/hep-th/9802150}
  {hep-th/9802150} \BibitemShut {NoStop}%
\bibitem [{\citenamefont {{Hayden}}\ and\ \citenamefont
  {{Preskill}}(2007)}]{HaydenPreskill}%
  \BibitemOpen
  \bibfield  {author} {\bibinfo {author} {\bibfnamefont {P.}~\bibnamefont
  {{Hayden}}}\ and\ \bibinfo {author} {\bibfnamefont {J.}~\bibnamefont
  {{Preskill}}},\ }\bibfield  {title} {\enquote {\bibinfo {title} {{Black holes
  as mirrors: quantum information in random subsystems}},}\ }\href {\doibase
  10.1088/1126-6708/2007/09/120} {\bibfield  {journal} {\bibinfo  {journal}
  {Journal of High Energy Physics}\ }\textbf {\bibinfo {volume} {9}},\ \bibinfo
  {eid} {120} (\bibinfo {year} {2007})},\ \Eprint
  {http://arxiv.org/abs/0708.4025} {arXiv:0708.4025 [hep-th]} \BibitemShut
  {NoStop}%
\bibitem [{\citenamefont {{Sekino}}\ and\ \citenamefont
  {{Susskind}}(2008)}]{SekinoSusskind}%
  \BibitemOpen
  \bibfield  {author} {\bibinfo {author} {\bibfnamefont {Y.}~\bibnamefont
  {{Sekino}}}\ and\ \bibinfo {author} {\bibfnamefont {L.}~\bibnamefont
  {{Susskind}}},\ }\bibfield  {title} {\enquote {\bibinfo {title} {{Fast
  scramblers}},}\ }\href {\doibase 10.1088/1126-6708/2008/10/065} {\bibfield
  {journal} {\bibinfo  {journal} {Journal of High Energy Physics}\ }\textbf
  {\bibinfo {volume} {10}},\ \bibinfo {eid} {065} (\bibinfo {year} {2008})},\
  \Eprint {http://arxiv.org/abs/0808.2096} {arXiv:0808.2096 [hep-th]}
  \BibitemShut {NoStop}%
\bibitem [{\citenamefont {{Hosur}}\ \emph {et~al.}(2016)\citenamefont
  {{Hosur}}, \citenamefont {{Qi}}, \citenamefont {{Roberts}},\ and\
  \citenamefont {{Yoshida}}}]{HosurYoshida}%
  \BibitemOpen
  \bibfield  {author} {\bibinfo {author} {\bibfnamefont {P.}~\bibnamefont
  {{Hosur}}}, \bibinfo {author} {\bibfnamefont {X.-L.}\ \bibnamefont {{Qi}}},
  \bibinfo {author} {\bibfnamefont {D.~A.}\ \bibnamefont {{Roberts}}}, \ and\
  \bibinfo {author} {\bibfnamefont {B.}~\bibnamefont {{Yoshida}}},\ }\bibfield
  {title} {\enquote {\bibinfo {title} {{Chaos in quantum channels}},}\ }\href
  {\doibase 10.1007/JHEP02(2016)004} {\bibfield  {journal} {\bibinfo  {journal}
  {Journal of High Energy Physics}\ }\textbf {\bibinfo {volume} {2}},\ \bibinfo
  {eid} {4} (\bibinfo {year} {2016})},\ \Eprint
  {http://arxiv.org/abs/1511.04021} {arXiv:1511.04021 [hep-th]} \BibitemShut
  {NoStop}%
\bibitem [{\citenamefont {{Shenker}}\ and\ \citenamefont
  {{Stanford}}(2014)}]{ShenkerStanfordButterfly}%
  \BibitemOpen
  \bibfield  {author} {\bibinfo {author} {\bibfnamefont {S.~H.}\ \bibnamefont
  {{Shenker}}}\ and\ \bibinfo {author} {\bibfnamefont {D.}~\bibnamefont
  {{Stanford}}},\ }\bibfield  {title} {\enquote {\bibinfo {title} {{Black holes
  and the butterfly effect}},}\ }\href {\doibase 10.1007/JHEP03(2014)067}
  {\bibfield  {journal} {\bibinfo  {journal} {Journal of High Energy Physics}\
  }\textbf {\bibinfo {volume} {3}},\ \bibinfo {eid} {67} (\bibinfo {year}
  {2014})},\ \Eprint {http://arxiv.org/abs/1306.0622} {arXiv:1306.0622
  [hep-th]} \BibitemShut {NoStop}%
\bibitem [{\citenamefont {{Lashkari}}\ \emph {et~al.}(2013)\citenamefont
  {{Lashkari}}, \citenamefont {{Stanford}}, \citenamefont {{Hastings}},
  \citenamefont {{Osborne}},\ and\ \citenamefont {{Hayden}}}]{Lashkari}%
  \BibitemOpen
  \bibfield  {author} {\bibinfo {author} {\bibfnamefont {N.}~\bibnamefont
  {{Lashkari}}}, \bibinfo {author} {\bibfnamefont {D.}~\bibnamefont
  {{Stanford}}}, \bibinfo {author} {\bibfnamefont {M.}~\bibnamefont
  {{Hastings}}}, \bibinfo {author} {\bibfnamefont {T.}~\bibnamefont
  {{Osborne}}}, \ and\ \bibinfo {author} {\bibfnamefont {P.}~\bibnamefont
  {{Hayden}}},\ }\bibfield  {title} {\enquote {\bibinfo {title} {{Towards the
  fast scrambling conjecture}},}\ }\href {\doibase 10.1007/JHEP04(2013)022}
  {\bibfield  {journal} {\bibinfo  {journal} {Journal of High Energy Physics}\
  }\textbf {\bibinfo {volume} {4}},\ \bibinfo {eid} {22} (\bibinfo {year}
  {2013})},\ \Eprint {http://arxiv.org/abs/1111.6580} {arXiv:1111.6580
  [hep-th]} \BibitemShut {NoStop}%
\bibitem [{\citenamefont {{Roberts}}\ \emph {et~al.}(2015)\citenamefont
  {{Roberts}}, \citenamefont {{Stanford}},\ and\ \citenamefont
  {{Susskind}}}]{LocalizedShocks}%
  \BibitemOpen
  \bibfield  {author} {\bibinfo {author} {\bibfnamefont {D.~A.}\ \bibnamefont
  {{Roberts}}}, \bibinfo {author} {\bibfnamefont {D.}~\bibnamefont
  {{Stanford}}}, \ and\ \bibinfo {author} {\bibfnamefont {L.}~\bibnamefont
  {{Susskind}}},\ }\bibfield  {title} {\enquote {\bibinfo {title} {{Localized
  shocks}},}\ }\href {\doibase 10.1007/JHEP03(2015)051} {\bibfield  {journal}
  {\bibinfo  {journal} {Journal of High Energy Physics}\ }\textbf {\bibinfo
  {volume} {3}},\ \bibinfo {eid} {51} (\bibinfo {year} {2015})},\ \Eprint
  {http://arxiv.org/abs/1409.8180} {arXiv:1409.8180 [hep-th]} \BibitemShut
  {NoStop}%
\bibitem [{\citenamefont {{Cotler}}\ \emph {et~al.}(2017)\citenamefont
  {{Cotler}}, \citenamefont {{Gur-Ari}}, \citenamefont {{Hanada}},
  \citenamefont {{Polchinski}}, \citenamefont {{Saad}}, \citenamefont
  {{Shenker}}, \citenamefont {{Stanford}}, \citenamefont {{Streicher}},\ and\
  \citenamefont {{Tezuka}}}]{CotlerRM}%
  \BibitemOpen
  \bibfield  {author} {\bibinfo {author} {\bibfnamefont {J.~S.}\ \bibnamefont
  {{Cotler}}}, \bibinfo {author} {\bibfnamefont {G.}~\bibnamefont {{Gur-Ari}}},
  \bibinfo {author} {\bibfnamefont {M.}~\bibnamefont {{Hanada}}}, \bibinfo
  {author} {\bibfnamefont {J.}~\bibnamefont {{Polchinski}}}, \bibinfo {author}
  {\bibfnamefont {P.}~\bibnamefont {{Saad}}}, \bibinfo {author} {\bibfnamefont
  {S.~H.}\ \bibnamefont {{Shenker}}}, \bibinfo {author} {\bibfnamefont
  {D.}~\bibnamefont {{Stanford}}}, \bibinfo {author} {\bibfnamefont
  {A.}~\bibnamefont {{Streicher}}}, \ and\ \bibinfo {author} {\bibfnamefont
  {M.}~\bibnamefont {{Tezuka}}},\ }\bibfield  {title} {\enquote {\bibinfo
  {title} {{Black holes and random matrices}},}\ }\href {\doibase
  10.1007/JHEP05(2017)118} {\bibfield  {journal} {\bibinfo  {journal} {Journal
  of High Energy Physics}\ }\textbf {\bibinfo {volume} {5}},\ \bibinfo {eid}
  {118} (\bibinfo {year} {2017})},\ \Eprint {http://arxiv.org/abs/1611.04650}
  {arXiv:1611.04650 [hep-th]} \BibitemShut {NoStop}%
\bibitem [{\citenamefont {Roberts}\ and\ \citenamefont
  {Stanford}(2015)}]{RobertsStanford}%
  \BibitemOpen
  \bibfield  {author} {\bibinfo {author} {\bibfnamefont {Daniel~A.}\
  \bibnamefont {Roberts}}\ and\ \bibinfo {author} {\bibfnamefont {Douglas}\
  \bibnamefont {Stanford}},\ }\bibfield  {title} {\enquote {\bibinfo {title}
  {Diagnosing chaos using four-point functions in two-dimensional conformal
  field theory},}\ }\href {\doibase 10.1103/PhysRevLett.115.131603} {\bibfield
  {journal} {\bibinfo  {journal} {Phys. Rev. Lett.}\ }\textbf {\bibinfo
  {volume} {115}},\ \bibinfo {pages} {131603} (\bibinfo {year}
  {2015})}\BibitemShut {NoStop}%
\bibitem [{\citenamefont {{Kitaev}}()}]{KitaevSYK}%
  \BibitemOpen
  \bibfield  {author} {\bibinfo {author} {\bibfnamefont {A}~\bibnamefont
  {{Kitaev}}},\ }\bibfield  {title} {\enquote {\bibinfo {title} {{A simple
  model of quantum holography}},}\ }\href {\doibase
  http://online.kitp.ucsb.edu/online/entangled15/kitaev/,http:
  //online.kitp.ucsb.edu/online/entangled15/kitaev2/} {\bibfield  {journal}
  {\bibinfo  {journal} {Talks at KITP, April 7, 2015 and May 27, 2015.}\
  }http://online.kitp.ucsb.edu/online/entangled15/kitaev/,http:
  //online.kitp.ucsb.edu/online/entangled15/kitaev2/}\BibitemShut {NoStop}%
\bibitem [{\citenamefont {{Sachdev}}\ and\ \citenamefont
  {{Ye}}(1993)}]{SachdevSYK}%
  \BibitemOpen
  \bibfield  {author} {\bibinfo {author} {\bibfnamefont {S.}~\bibnamefont
  {{Sachdev}}}\ and\ \bibinfo {author} {\bibfnamefont {J.}~\bibnamefont
  {{Ye}}},\ }\bibfield  {title} {\enquote {\bibinfo {title} {{Gapless
  spin-fluid ground state in a random quantum Heisenberg magnet}},}\ }\href
  {\doibase 10.1103/PhysRevLett.70.3339} {\bibfield  {journal} {\bibinfo
  {journal} {Physical Review Letters}\ }\textbf {\bibinfo {volume} {70}},\
  \bibinfo {pages} {3339--3342} (\bibinfo {year} {1993})},\ \Eprint
  {http://arxiv.org/abs/cond-mat/9212030} {cond-mat/9212030} \BibitemShut
  {NoStop}%
\bibitem [{\citenamefont {Lieb}\ and\ \citenamefont {Robinson}()}]{Lieb72}%
  \BibitemOpen
  \bibfield  {author} {\bibinfo {author} {\bibfnamefont {Elliott~H.}\
  \bibnamefont {Lieb}}\ and\ \bibinfo {author} {\bibfnamefont {Derek~W.}\
  \bibnamefont {Robinson}},\ }\bibfield  {title} {\enquote {\bibinfo {title}
  {The finite group velocity of quantum spin systems},}\ }\href {\doibase
  10.1007/BF01645779} {\bibfield  {journal} {\bibinfo  {journal}
  {Communications in Mathematical Physics}\ }\textbf {\bibinfo {volume} {28}},\
  \bibinfo {pages} {251--257}}\BibitemShut {NoStop}%
\bibitem [{\citenamefont {{Larkin}}\ and\ \citenamefont
  {{Ovchinnikov}}(1969)}]{Larkinotoc}%
  \BibitemOpen
  \bibfield  {author} {\bibinfo {author} {\bibfnamefont {A.~I.}\ \bibnamefont
  {{Larkin}}}\ and\ \bibinfo {author} {\bibfnamefont {Y.~N.}\ \bibnamefont
  {{Ovchinnikov}}},\ }\bibfield  {title} {\enquote {\bibinfo {title}
  {{Quasiclassical Method in the Theory of Superconductivity}},}\ }\href@noop
  {} {\bibfield  {journal} {\bibinfo  {journal} {Soviet Journal of Experimental
  and Theoretical Physics}\ }\textbf {\bibinfo {volume} {28}},\ \bibinfo
  {pages} {1200} (\bibinfo {year} {1969})}\BibitemShut {NoStop}%
\bibitem [{\citenamefont {{Maldacena}}\ \emph {et~al.}(2016)\citenamefont
  {{Maldacena}}, \citenamefont {{Shenker}},\ and\ \citenamefont
  {{Stanford}}}]{chaosbound}%
  \BibitemOpen
  \bibfield  {author} {\bibinfo {author} {\bibfnamefont {J.}~\bibnamefont
  {{Maldacena}}}, \bibinfo {author} {\bibfnamefont {S.~H.}\ \bibnamefont
  {{Shenker}}}, \ and\ \bibinfo {author} {\bibfnamefont {D.}~\bibnamefont
  {{Stanford}}},\ }\bibfield  {title} {\enquote {\bibinfo {title} {{A bound on
  chaos}},}\ }\href {\doibase 10.1007/JHEP08(2016)106} {\bibfield  {journal}
  {\bibinfo  {journal} {Journal of High Energy Physics}\ }\textbf {\bibinfo
  {volume} {8}},\ \bibinfo {eid} {106} (\bibinfo {year} {2016})},\ \Eprint
  {http://arxiv.org/abs/1503.01409} {arXiv:1503.01409 [hep-th]} \BibitemShut
  {NoStop}%
\bibitem [{\citenamefont {{Gu}}\ \emph {et~al.}(2017)\citenamefont {{Gu}},
  \citenamefont {{Qi}},\ and\ \citenamefont {{Stanford}}}]{GuQiStanford}%
  \BibitemOpen
  \bibfield  {author} {\bibinfo {author} {\bibfnamefont {Y.}~\bibnamefont
  {{Gu}}}, \bibinfo {author} {\bibfnamefont {X.-L.}\ \bibnamefont {{Qi}}}, \
  and\ \bibinfo {author} {\bibfnamefont {D.}~\bibnamefont {{Stanford}}},\
  }\bibfield  {title} {\enquote {\bibinfo {title} {{Local criticality,
  diffusion and chaos in generalized Sachdev-Ye-Kitaev models}},}\ }\href
  {\doibase 10.1007/JHEP05(2017)125} {\bibfield  {journal} {\bibinfo  {journal}
  {Journal of High Energy Physics}\ }\textbf {\bibinfo {volume} {5}},\ \bibinfo
  {eid} {125} (\bibinfo {year} {2017})},\ \Eprint
  {http://arxiv.org/abs/1609.07832} {arXiv:1609.07832 [hep-th]} \BibitemShut
  {NoStop}%
\bibitem [{\citenamefont {{Gu}}\ and\ \citenamefont {{Qi}}(2016)}]{GuQi_rcft}%
  \BibitemOpen
  \bibfield  {author} {\bibinfo {author} {\bibfnamefont {Y.}~\bibnamefont
  {{Gu}}}\ and\ \bibinfo {author} {\bibfnamefont {X.-L.}\ \bibnamefont
  {{Qi}}},\ }\bibfield  {title} {\enquote {\bibinfo {title} {{Fractional
  statistics and the butterfly effect}},}\ }\href {\doibase
  10.1007/JHEP08(2016)129} {\bibfield  {journal} {\bibinfo  {journal} {Journal
  of High Energy Physics}\ }\textbf {\bibinfo {volume} {8}},\ \bibinfo {eid}
  {129} (\bibinfo {year} {2016})},\ \Eprint {http://arxiv.org/abs/1602.06543}
  {arXiv:1602.06543 [hep-th]} \BibitemShut {NoStop}%
\bibitem [{\citenamefont {{Stanford}}(2016)}]{StanfordWeakCoupling}%
  \BibitemOpen
  \bibfield  {author} {\bibinfo {author} {\bibfnamefont {D.}~\bibnamefont
  {{Stanford}}},\ }\bibfield  {title} {\enquote {\bibinfo {title} {{Many-body
  chaos at weak coupling}},}\ }\href {\doibase 10.1007/JHEP10(2016)009}
  {\bibfield  {journal} {\bibinfo  {journal} {Journal of High Energy Physics}\
  }\textbf {\bibinfo {volume} {10}},\ \bibinfo {eid} {9} (\bibinfo {year}
  {2016})},\ \Eprint {http://arxiv.org/abs/1512.07687} {arXiv:1512.07687
  [hep-th]} \BibitemShut {NoStop}%
\bibitem [{\citenamefont {{Patel}}\ \emph {et~al.}(2017)\citenamefont
  {{Patel}}, \citenamefont {{Chowdhury}}, \citenamefont {{Sachdev}},\ and\
  \citenamefont {{Swingle}}}]{PatelDiffusiveMetal}%
  \BibitemOpen
  \bibfield  {author} {\bibinfo {author} {\bibfnamefont {A.~A.}\ \bibnamefont
  {{Patel}}}, \bibinfo {author} {\bibfnamefont {D.}~\bibnamefont
  {{Chowdhury}}}, \bibinfo {author} {\bibfnamefont {S.}~\bibnamefont
  {{Sachdev}}}, \ and\ \bibinfo {author} {\bibfnamefont {B.}~\bibnamefont
  {{Swingle}}},\ }\bibfield  {title} {\enquote {\bibinfo {title} {{Quantum
  Butterfly Effect in Weakly Interacting Diffusive Metals}},}\ }\href {\doibase
  10.1103/PhysRevX.7.031047} {\bibfield  {journal} {\bibinfo  {journal}
  {Physical Review X}\ }\textbf {\bibinfo {volume} {7}},\ \bibinfo {eid}
  {031047} (\bibinfo {year} {2017})},\ \Eprint
  {http://arxiv.org/abs/1703.07353} {arXiv:1703.07353 [cond-mat.str-el]}
  \BibitemShut {NoStop}%
\bibitem [{\citenamefont {{Chowdhury}}\ and\ \citenamefont
  {{Swingle}}(2017)}]{ChowdhuryON}%
  \BibitemOpen
  \bibfield  {author} {\bibinfo {author} {\bibfnamefont {D.}~\bibnamefont
  {{Chowdhury}}}\ and\ \bibinfo {author} {\bibfnamefont {B.}~\bibnamefont
  {{Swingle}}},\ }\bibfield  {title} {\enquote {\bibinfo {title} {{Onset of
  many-body chaos in the O (N ) model}},}\ }\href {\doibase
  10.1103/PhysRevD.96.065005} {\bibfield  {journal} {\bibinfo  {journal}
  {\prd}\ }\textbf {\bibinfo {volume} {96}},\ \bibinfo {eid} {065005} (\bibinfo
  {year} {2017})},\ \Eprint {http://arxiv.org/abs/1703.02545} {arXiv:1703.02545
  [cond-mat.str-el]} \BibitemShut {NoStop}%
\bibitem [{\citenamefont {{Rozenbaum}}\ \emph {et~al.}(2017)\citenamefont
  {{Rozenbaum}}, \citenamefont {{Ganeshan}},\ and\ \citenamefont
  {{Galitski}}}]{Galitski_lyapunov}%
  \BibitemOpen
  \bibfield  {author} {\bibinfo {author} {\bibfnamefont {E.~B.}\ \bibnamefont
  {{Rozenbaum}}}, \bibinfo {author} {\bibfnamefont {S.}~\bibnamefont
  {{Ganeshan}}}, \ and\ \bibinfo {author} {\bibfnamefont {V.}~\bibnamefont
  {{Galitski}}},\ }\bibfield  {title} {\enquote {\bibinfo {title} {{Lyapunov
  Exponent and Out-of-Time-Ordered Correlator's Growth Rate in a Chaotic
  System}},}\ }\href {\doibase 10.1103/PhysRevLett.118.086801} {\bibfield
  {journal} {\bibinfo  {journal} {Physical Review Letters}\ }\textbf {\bibinfo
  {volume} {118}},\ \bibinfo {eid} {086801} (\bibinfo {year} {2017})},\ \Eprint
  {http://arxiv.org/abs/1609.01707} {arXiv:1609.01707 [cond-mat.dis-nn]}
  \BibitemShut {NoStop}%
\bibitem [{\citenamefont {{D{\'o}ra}}\ and\ \citenamefont
  {{Moessner}}(2017)}]{DoraMoessner}%
  \BibitemOpen
  \bibfield  {author} {\bibinfo {author} {\bibfnamefont {B.}~\bibnamefont
  {{D{\'o}ra}}}\ and\ \bibinfo {author} {\bibfnamefont {R.}~\bibnamefont
  {{Moessner}}},\ }\bibfield  {title} {\enquote {\bibinfo {title}
  {{Out-of-Time-Ordered Density Correlators in Luttinger Liquids}},}\ }\href
  {\doibase 10.1103/PhysRevLett.119.026802} {\bibfield  {journal} {\bibinfo
  {journal} {Physical Review Letters}\ }\textbf {\bibinfo {volume} {119}},\
  \bibinfo {eid} {026802} (\bibinfo {year} {2017})},\ \Eprint
  {http://arxiv.org/abs/1612.00614} {arXiv:1612.00614 [cond-mat.str-el]}
  \BibitemShut {NoStop}%
\bibitem [{\citenamefont {{Luitz}}\ and\ \citenamefont {{Bar
  Lev}}(2017)}]{LuitzScrambling}%
  \BibitemOpen
  \bibfield  {author} {\bibinfo {author} {\bibfnamefont {D.~J.}\ \bibnamefont
  {{Luitz}}}\ and\ \bibinfo {author} {\bibfnamefont {Y.}~\bibnamefont {{Bar
  Lev}}},\ }\bibfield  {title} {\enquote {\bibinfo {title} {{Information
  propagation in isolated quantum systems}},}\ }\href {\doibase
  10.1103/PhysRevB.96.020406} {\bibfield  {journal} {\bibinfo  {journal}
  {\prb}\ }\textbf {\bibinfo {volume} {96}},\ \bibinfo {eid} {020406} (\bibinfo
  {year} {2017})},\ \Eprint {http://arxiv.org/abs/1702.03929} {arXiv:1702.03929
  [cond-mat.dis-nn]} \BibitemShut {NoStop}%
\bibitem [{\citenamefont {{Kukuljan}}\ \emph {et~al.}(2017)\citenamefont
  {{Kukuljan}}, \citenamefont {{Grozdanov}},\ and\ \citenamefont
  {{Prosen}}}]{ProsenWeakChaos}%
  \BibitemOpen
  \bibfield  {author} {\bibinfo {author} {\bibfnamefont {I.}~\bibnamefont
  {{Kukuljan}}}, \bibinfo {author} {\bibfnamefont {S.}~\bibnamefont
  {{Grozdanov}}}, \ and\ \bibinfo {author} {\bibfnamefont {T.}~\bibnamefont
  {{Prosen}}},\ }\bibfield  {title} {\enquote {\bibinfo {title} {{Weak quantum
  chaos}},}\ }\href {\doibase 10.1103/PhysRevB.96.060301} {\bibfield  {journal}
  {\bibinfo  {journal} {\prb}\ }\textbf {\bibinfo {volume} {96}},\ \bibinfo
  {eid} {060301} (\bibinfo {year} {2017})},\ \Eprint
  {http://arxiv.org/abs/1701.09147} {arXiv:1701.09147 [cond-mat.stat-mech]}
  \BibitemShut {NoStop}%
\bibitem [{\citenamefont {{Aleiner}}\ \emph {et~al.}(2016)\citenamefont
  {{Aleiner}}, \citenamefont {{Faoro}},\ and\ \citenamefont
  {{Ioffe}}}]{AleinerOTOC}%
  \BibitemOpen
  \bibfield  {author} {\bibinfo {author} {\bibfnamefont {I.~L.}\ \bibnamefont
  {{Aleiner}}}, \bibinfo {author} {\bibfnamefont {L.}~\bibnamefont {{Faoro}}},
  \ and\ \bibinfo {author} {\bibfnamefont {L.~B.}\ \bibnamefont {{Ioffe}}},\
  }\bibfield  {title} {\enquote {\bibinfo {title} {{Microscopic model of
  quantum butterfly effect: Out-of-time-order correlators and traveling
  combustion waves}},}\ }\href {\doibase 10.1016/j.aop.2016.09.006} {\bibfield
  {journal} {\bibinfo  {journal} {Annals of Physics}\ }\textbf {\bibinfo
  {volume} {375}},\ \bibinfo {pages} {378--406} (\bibinfo {year} {2016})},\
  \Eprint {http://arxiv.org/abs/1609.01251} {arXiv:1609.01251
  [cond-mat.stat-mech]} \BibitemShut {NoStop}%
\bibitem [{\citenamefont {{Lin}}\ and\ \citenamefont
  {{Motrunich}}(2018)}]{MotrunichTFIM_otoc}%
  \BibitemOpen
  \bibfield  {author} {\bibinfo {author} {\bibfnamefont {C.-J.}\ \bibnamefont
  {{Lin}}}\ and\ \bibinfo {author} {\bibfnamefont {O.~I.}\ \bibnamefont
  {{Motrunich}}},\ }\bibfield  {title} {\enquote {\bibinfo {title}
  {{Out-of-time-ordered correlators in quantum Ising chain}},}\ }\href@noop {}
  {\bibfield  {journal} {\bibinfo  {journal} {ArXiv e-prints}\ } (\bibinfo
  {year} {2018})},\ \Eprint {http://arxiv.org/abs/1801.01636} {arXiv:1801.01636
  [cond-mat.stat-mech]} \BibitemShut {NoStop}%
\bibitem [{\citenamefont {{Chen}}\ \emph {et~al.}(2017)\citenamefont {{Chen}},
  \citenamefont {{Zhou}}, \citenamefont {{Huse}},\ and\ \citenamefont
  {{Fradkin}}}]{FradkinHuse}%
  \BibitemOpen
  \bibfield  {author} {\bibinfo {author} {\bibfnamefont {X.}~\bibnamefont
  {{Chen}}}, \bibinfo {author} {\bibfnamefont {T.}~\bibnamefont {{Zhou}}},
  \bibinfo {author} {\bibfnamefont {D.~A.}\ \bibnamefont {{Huse}}}, \ and\
  \bibinfo {author} {\bibfnamefont {E.}~\bibnamefont {{Fradkin}}},\ }\bibfield
  {title} {\enquote {\bibinfo {title} {{Out-of-time-order correlations in
  many-body localized and thermal phases}},}\ }\href {\doibase
  10.1002/andp.201600332} {\bibfield  {journal} {\bibinfo  {journal} {Annalen
  der Physik}\ }\textbf {\bibinfo {volume} {529}},\ \bibinfo {pages} {1600332}
  (\bibinfo {year} {2017})}\BibitemShut {NoStop}%
\bibitem [{\citenamefont {{Chan}}\ \emph {et~al.}(2017)\citenamefont {{Chan}},
  \citenamefont {{De Luca}},\ and\ \citenamefont
  {{Chalker}}}]{ChalkerFloquetChaos}%
  \BibitemOpen
  \bibfield  {author} {\bibinfo {author} {\bibfnamefont {A.}~\bibnamefont
  {{Chan}}}, \bibinfo {author} {\bibfnamefont {A.}~\bibnamefont {{De Luca}}}, \
  and\ \bibinfo {author} {\bibfnamefont {J.~T.}\ \bibnamefont {{Chalker}}},\
  }\bibfield  {title} {\enquote {\bibinfo {title} {{Solution of a minimal model
  for many-body quantum chaos}},}\ }\href@noop {} {\bibfield  {journal}
  {\bibinfo  {journal} {ArXiv e-prints}\ } (\bibinfo {year} {2017})},\ \Eprint
  {http://arxiv.org/abs/1712.06836} {arXiv:1712.06836 [cond-mat.stat-mech]}
  \BibitemShut {NoStop}%
\bibitem [{\citenamefont {{Brown}}\ and\ \citenamefont
  {{Fawzi}}(2012)}]{FawziScrambling}%
  \BibitemOpen
  \bibfield  {author} {\bibinfo {author} {\bibfnamefont {W.}~\bibnamefont
  {{Brown}}}\ and\ \bibinfo {author} {\bibfnamefont {O.}~\bibnamefont
  {{Fawzi}}},\ }\bibfield  {title} {\enquote {\bibinfo {title} {{Scrambling
  speed of random quantum circuits}},}\ }\href@noop {} {\bibfield  {journal}
  {\bibinfo  {journal} {ArXiv e-prints}\ } (\bibinfo {year} {2012})},\ \Eprint
  {http://arxiv.org/abs/1210.6644} {arXiv:1210.6644 [quant-ph]} \BibitemShut
  {NoStop}%
\bibitem [{\citenamefont {{Nahum}}\ \emph
  {et~al.}(2017{\natexlab{a}})\citenamefont {{Nahum}}, \citenamefont
  {{Vijay}},\ and\ \citenamefont {{Haah}}}]{opspreadAdam}%
  \BibitemOpen
  \bibfield  {author} {\bibinfo {author} {\bibfnamefont {A.}~\bibnamefont
  {{Nahum}}}, \bibinfo {author} {\bibfnamefont {S.}~\bibnamefont {{Vijay}}}, \
  and\ \bibinfo {author} {\bibfnamefont {J.}~\bibnamefont {{Haah}}},\
  }\bibfield  {title} {\enquote {\bibinfo {title} {{Operator Spreading in
  Random Unitary Circuits}},}\ }\href@noop {} {\bibfield  {journal} {\bibinfo
  {journal} {ArXiv e-prints}\ } (\bibinfo {year} {2017}{\natexlab{a}})},\
  \Eprint {http://arxiv.org/abs/1705.08975} {arXiv:1705.08975
  [cond-mat.str-el]} \BibitemShut {NoStop}%
\bibitem [{\citenamefont {{von Keyserlingk}}\ \emph {et~al.}(2017)\citenamefont
  {{von Keyserlingk}}, \citenamefont {{Rakovszky}}, \citenamefont
  {{Pollmann}},\ and\ \citenamefont {{Sondhi}}}]{opspreadCurt}%
  \BibitemOpen
  \bibfield  {author} {\bibinfo {author} {\bibfnamefont {C.W.}\ \bibnamefont
  {{von Keyserlingk}}}, \bibinfo {author} {\bibfnamefont {T.}~\bibnamefont
  {{Rakovszky}}}, \bibinfo {author} {\bibfnamefont {F.}~\bibnamefont
  {{Pollmann}}}, \ and\ \bibinfo {author} {\bibfnamefont {S.}~\bibnamefont
  {{Sondhi}}},\ }\bibfield  {title} {\enquote {\bibinfo {title} {{Operator
  hydrodynamics, OTOCs, and entanglement growth in systems without conservation
  laws}},}\ }\href@noop {} {\bibfield  {journal} {\bibinfo  {journal} {ArXiv
  e-prints}\ } (\bibinfo {year} {2017})},\ \Eprint
  {http://arxiv.org/abs/1705.08910} {arXiv:1705.08910 [cond-mat.str-el]}
  \BibitemShut {NoStop}%
\bibitem [{\citenamefont {{Rakovszky}}\ \emph {et~al.}(2017)\citenamefont
  {{Rakovszky}}, \citenamefont {{Pollmann}},\ and\ \citenamefont {{von
  Keyserlingk}}}]{TiborCons}%
  \BibitemOpen
  \bibfield  {author} {\bibinfo {author} {\bibfnamefont {T.}~\bibnamefont
  {{Rakovszky}}}, \bibinfo {author} {\bibfnamefont {F.}~\bibnamefont
  {{Pollmann}}}, \ and\ \bibinfo {author} {\bibfnamefont {C.~W.}\ \bibnamefont
  {{von Keyserlingk}}},\ }\bibfield  {title} {\enquote {\bibinfo {title}
  {{Diffusive hydrodynamics of out-of-time-ordered correlators with charge
  conservation}},}\ }\href@noop {} {\bibfield  {journal} {\bibinfo  {journal}
  {ArXiv e-prints}\ } (\bibinfo {year} {2017})},\ \Eprint
  {http://arxiv.org/abs/1710.09827} {arXiv:1710.09827 [cond-mat.stat-mech]}
  \BibitemShut {NoStop}%
\bibitem [{\citenamefont {{Khemani}}\ \emph {et~al.}(2017)\citenamefont
  {{Khemani}}, \citenamefont {{Vishwanath}},\ and\ \citenamefont
  {{Huse}}}]{KhemaniCons}%
  \BibitemOpen
  \bibfield  {author} {\bibinfo {author} {\bibfnamefont {V.}~\bibnamefont
  {{Khemani}}}, \bibinfo {author} {\bibfnamefont {A.}~\bibnamefont
  {{Vishwanath}}}, \ and\ \bibinfo {author} {\bibfnamefont {D.~A.}\
  \bibnamefont {{Huse}}},\ }\bibfield  {title} {\enquote {\bibinfo {title}
  {{Operator spreading and the emergence of dissipation in unitary dynamics
  with conservation laws}},}\ }\href@noop {} {\bibfield  {journal} {\bibinfo
  {journal} {ArXiv e-prints}\ } (\bibinfo {year} {2017})},\ \Eprint
  {http://arxiv.org/abs/1710.09835} {arXiv:1710.09835 [cond-mat.stat-mech]}
  \BibitemShut {NoStop}%
\bibitem [{\citenamefont {Bardarson}\ \emph {et~al.}(2012)\citenamefont
  {Bardarson}, \citenamefont {Pollmann},\ and\ \citenamefont
  {Moore}}]{BardarsonPollmannMoore}%
  \BibitemOpen
  \bibfield  {author} {\bibinfo {author} {\bibfnamefont {Jens~H.}\ \bibnamefont
  {Bardarson}}, \bibinfo {author} {\bibfnamefont {Frank}\ \bibnamefont
  {Pollmann}}, \ and\ \bibinfo {author} {\bibfnamefont {Joel~E.}\ \bibnamefont
  {Moore}},\ }\bibfield  {title} {\enquote {\bibinfo {title} {Unbounded growth
  of entanglement in models of many-body localization},}\ }\href {\doibase
  10.1103/PhysRevLett.109.017202} {\bibfield  {journal} {\bibinfo  {journal}
  {Phys. Rev. Lett.}\ }\textbf {\bibinfo {volume} {109}},\ \bibinfo {pages}
  {017202} (\bibinfo {year} {2012})}\BibitemShut {NoStop}%
\bibitem [{\citenamefont {Potter}\ \emph {et~al.}(2015)\citenamefont {Potter},
  \citenamefont {Vasseur},\ and\ \citenamefont {Parameswaran}}]{PVP}%
  \BibitemOpen
  \bibfield  {author} {\bibinfo {author} {\bibfnamefont {Andrew~C.}\
  \bibnamefont {Potter}}, \bibinfo {author} {\bibfnamefont {Romain}\
  \bibnamefont {Vasseur}}, \ and\ \bibinfo {author} {\bibfnamefont {S.~A.}\
  \bibnamefont {Parameswaran}},\ }\bibfield  {title} {\enquote {\bibinfo
  {title} {Universal properties of many-body delocalization transitions},}\
  }\href {\doibase 10.1103/PhysRevX.5.031033} {\bibfield  {journal} {\bibinfo
  {journal} {Phys. Rev. X}\ }\textbf {\bibinfo {volume} {5}},\ \bibinfo {pages}
  {031033} (\bibinfo {year} {2015})}\BibitemShut {NoStop}%
\bibitem [{\citenamefont {Vosk}\ \emph {et~al.}(2015)\citenamefont {Vosk},
  \citenamefont {Huse},\ and\ \citenamefont {Altman}}]{VHA}%
  \BibitemOpen
  \bibfield  {author} {\bibinfo {author} {\bibfnamefont {Ronen}\ \bibnamefont
  {Vosk}}, \bibinfo {author} {\bibfnamefont {David~A.}\ \bibnamefont {Huse}}, \
  and\ \bibinfo {author} {\bibfnamefont {Ehud}\ \bibnamefont {Altman}},\
  }\bibfield  {title} {\enquote {\bibinfo {title} {Theory of the many-body
  localization transition in one-dimensional systems},}\ }\href {\doibase
  10.1103/PhysRevX.5.031032} {\bibfield  {journal} {\bibinfo  {journal} {Phys.
  Rev. X}\ }\textbf {\bibinfo {volume} {5}},\ \bibinfo {pages} {031032}
  (\bibinfo {year} {2015})}\BibitemShut {NoStop}%
\bibitem [{\citenamefont {{Nahum}}\ \emph
  {et~al.}(2017{\natexlab{b}})\citenamefont {{Nahum}}, \citenamefont
  {{Ruhman}},\ and\ \citenamefont {{Huse}}}]{NahumRuhmanHuse}%
  \BibitemOpen
  \bibfield  {author} {\bibinfo {author} {\bibfnamefont {A.}~\bibnamefont
  {{Nahum}}}, \bibinfo {author} {\bibfnamefont {J.}~\bibnamefont {{Ruhman}}}, \
  and\ \bibinfo {author} {\bibfnamefont {D.~A.}\ \bibnamefont {{Huse}}},\
  }\bibfield  {title} {\enquote {\bibinfo {title} {{Dynamics of entanglement
  and transport in 1D systems with quenched randomness}},}\ }\href@noop {}
  {\bibfield  {journal} {\bibinfo  {journal} {ArXiv e-prints}\ } (\bibinfo
  {year} {2017}{\natexlab{b}})},\ \Eprint {http://arxiv.org/abs/1705.10364}
  {arXiv:1705.10364 [cond-mat.dis-nn]} \BibitemShut {NoStop}%
\bibitem [{Note1()}]{Note1}%
  \BibitemOpen
  \bibinfo {note} {We note that the usual definition of the classical Lyapunov
  exponent involves averaging the logarithm of the factor by which
  perturbations grow over initial states and perturbations. This is subtly
  different from the classical analog of the quantum OTOC where the
  commutator/Poisson bracket is averaged before taking the logarithm. It is
  worth exploring in future studies whether or not this difference in
  definitions has any qualitative consequences\cite
  {Galitski_lyapunov}.}\BibitemShut {Stop}%
\bibitem [{\citenamefont {Deissler}(1984)}]{Deissler1984}%
  \BibitemOpen
  \bibfield  {author} {\bibinfo {author} {\bibfnamefont {Robert~J.}\
  \bibnamefont {Deissler}},\ }\bibfield  {title} {\enquote {\bibinfo {title}
  {One-dimensional strings, random fluctuations, and complex chaotic
  structures},}\ }\href {\doibase https://doi.org/10.1016/0375-9601(84)90823-5}
  {\bibfield  {journal} {\bibinfo  {journal} {Physics Letters A}\ }\textbf
  {\bibinfo {volume} {100}},\ \bibinfo {pages} {451 -- 454} (\bibinfo {year}
  {1984})}\BibitemShut {NoStop}%
\bibitem [{\citenamefont {Kaneko}(1986)}]{Kaneko1986}%
  \BibitemOpen
  \bibfield  {author} {\bibinfo {author} {\bibfnamefont {Kunihiko}\
  \bibnamefont {Kaneko}},\ }\bibfield  {title} {\enquote {\bibinfo {title}
  {Lyapunov analysis and information flow in coupled map lattices},}\ }\href
  {\doibase https://doi.org/10.1016/0167-2789(86)90149-1} {\bibfield  {journal}
  {\bibinfo  {journal} {Physica D: Nonlinear Phenomena}\ }\textbf {\bibinfo
  {volume} {23}},\ \bibinfo {pages} {436 -- 447} (\bibinfo {year}
  {1986})}\BibitemShut {NoStop}%
\bibitem [{\citenamefont {Deissler}\ and\ \citenamefont
  {Kaneko}(1987)}]{DeisslerKaneko}%
  \BibitemOpen
  \bibfield  {author} {\bibinfo {author} {\bibfnamefont {Robert~J.}\
  \bibnamefont {Deissler}}\ and\ \bibinfo {author} {\bibfnamefont {Kunihiko}\
  \bibnamefont {Kaneko}},\ }\bibfield  {title} {\enquote {\bibinfo {title}
  {Velocity-dependent lyapunov exponents as a measure of chaos for open-flow
  systems},}\ }\href {\doibase https://doi.org/10.1016/0375-9601(87)90581-0}
  {\bibfield  {journal} {\bibinfo  {journal} {Physics Letters A}\ }\textbf
  {\bibinfo {volume} {119}},\ \bibinfo {pages} {397 -- 402} (\bibinfo {year}
  {1987})}\BibitemShut {NoStop}%
\bibitem [{\citenamefont {{Calabrese}}\ and\ \citenamefont
  {{Cardy}}(2009)}]{CalabreseCardy}%
  \BibitemOpen
  \bibfield  {author} {\bibinfo {author} {\bibfnamefont {P.}~\bibnamefont
  {{Calabrese}}}\ and\ \bibinfo {author} {\bibfnamefont {J.}~\bibnamefont
  {{Cardy}}},\ }\bibfield  {title} {\enquote {\bibinfo {title} {{Entanglement
  entropy and conformal field theory}},}\ }\href {\doibase
  10.1088/1751-8113/42/50/504005} {\bibfield  {journal} {\bibinfo  {journal}
  {Journal of Physics A Mathematical General}\ }\textbf {\bibinfo {volume}
  {42}},\ \bibinfo {eid} {504005} (\bibinfo {year} {2009})},\ \Eprint
  {http://arxiv.org/abs/0905.4013} {arXiv:0905.4013 [cond-mat.stat-mech]}
  \BibitemShut {NoStop}%
\bibitem [{\citenamefont {Prosen}\ and\ \citenamefont
  {Pi{\v{z}}orn}(2007)}]{prosen2007operator}%
  \BibitemOpen
  \bibfield  {author} {\bibinfo {author} {\bibfnamefont {Toma{\v{z}}}\
  \bibnamefont {Prosen}}\ and\ \bibinfo {author} {\bibfnamefont {Iztok}\
  \bibnamefont {Pi{\v{z}}orn}},\ }\bibfield  {title} {\enquote {\bibinfo
  {title} {Operator space entanglement entropy in a transverse ising chain},}\
  }\href@noop {} {\bibfield  {journal} {\bibinfo  {journal} {Physical Review
  A}\ }\textbf {\bibinfo {volume} {76}},\ \bibinfo {pages} {032316} (\bibinfo
  {year} {2007})}\BibitemShut {NoStop}%
\bibitem [{\citenamefont {Pizorn}\ and\ \citenamefont
  {Prosen}(2009)}]{pizorn2009operator}%
  \BibitemOpen
  \bibfield  {author} {\bibinfo {author} {\bibfnamefont {Iztok}\ \bibnamefont
  {Pizorn}}\ and\ \bibinfo {author} {\bibfnamefont {Tomaz}\ \bibnamefont
  {Prosen}},\ }\bibfield  {title} {\enquote {\bibinfo {title} {Operator space
  entanglement entropy in xy spin chains},}\ }\href@noop {} {\bibfield
  {journal} {\bibinfo  {journal} {arXiv preprint arXiv:0903.2432}\ } (\bibinfo
  {year} {2009})}\BibitemShut {NoStop}%
\bibitem [{\citenamefont {Dubail}(2017)}]{dubail2017entanglement}%
  \BibitemOpen
  \bibfield  {author} {\bibinfo {author} {\bibfnamefont {J}~\bibnamefont
  {Dubail}},\ }\bibfield  {title} {\enquote {\bibinfo {title} {Entanglement
  scaling of operators: a conformal field theory approach, with a glimpse of
  simulability of long-time dynamics in 1+ 1d},}\ }\href@noop {} {\bibfield
  {journal} {\bibinfo  {journal} {Journal of Physics A: Mathematical and
  Theoretical}\ }\textbf {\bibinfo {volume} {50}},\ \bibinfo {pages} {234001}
  (\bibinfo {year} {2017})}\BibitemShut {NoStop}%
\bibitem [{\citenamefont {Jonay}\ \emph {et~al.}()\citenamefont {Jonay},
  \citenamefont {Huse},\ and\ \citenamefont {Nahum}}]{jonay}%
  \BibitemOpen
  \bibfield  {author} {\bibinfo {author} {\bibfnamefont {C.}~\bibnamefont
  {Jonay}}, \bibinfo {author} {\bibfnamefont {D.A.}\ \bibnamefont {Huse}}, \
  and\ \bibinfo {author} {\bibfnamefont {A.}~\bibnamefont {Nahum}},\
  }\href@noop {} {\bibfield  {journal} {\bibinfo  {journal} {Coarse-grained
  dynamics of operator and state entanglement}\ }}\Eprint
  {http://arxiv.org/abs/1803.00089} {1803.00089} \BibitemShut {NoStop}%
\bibitem [{\citenamefont {{Nachtergaele}}\ \emph {et~al.}(2006)\citenamefont
  {{Nachtergaele}}, \citenamefont {{Ogata}},\ and\ \citenamefont
  {{Sims}}}]{Nachtergaele1}%
  \BibitemOpen
  \bibfield  {author} {\bibinfo {author} {\bibfnamefont {B.}~\bibnamefont
  {{Nachtergaele}}}, \bibinfo {author} {\bibfnamefont {Y.}~\bibnamefont
  {{Ogata}}}, \ and\ \bibinfo {author} {\bibfnamefont {R.}~\bibnamefont
  {{Sims}}},\ }\bibfield  {title} {\enquote {\bibinfo {title} {{Propagation of
  Correlations in Quantum Lattice Systems}},}\ }\href@noop {} {\bibfield
  {journal} {\bibinfo  {journal} {Journal of Statistical Physics}\ }\textbf
  {\bibinfo {volume} {124}},\ \bibinfo {pages} {1--13} (\bibinfo {year}
  {2006})},\ \Eprint {http://arxiv.org/abs/math-ph/0603064} {math-ph/0603064}
  \BibitemShut {NoStop}%
\bibitem [{\citenamefont {{Nachtergaele}}\ and\ \citenamefont
  {{Sims}}(2006)}]{Nachtergaele2}%
  \BibitemOpen
  \bibfield  {author} {\bibinfo {author} {\bibfnamefont {B.}~\bibnamefont
  {{Nachtergaele}}}\ and\ \bibinfo {author} {\bibfnamefont {R.}~\bibnamefont
  {{Sims}}},\ }\bibfield  {title} {\enquote {\bibinfo {title} {{Lieb-Robinson
  Bounds and the Exponential Clustering Theorem}},}\ }\href@noop {} {\bibfield
  {journal} {\bibinfo  {journal} {Communications in Mathematical Physics}\
  }\textbf {\bibinfo {volume} {265}},\ \bibinfo {pages} {119--130} (\bibinfo
  {year} {2006})},\ \Eprint {http://arxiv.org/abs/math-ph/0506030}
  {math-ph/0506030} \BibitemShut {NoStop}%
\bibitem [{\citenamefont {{Hastings}}\ and\ \citenamefont
  {{Koma}}(2006)}]{HastingsSpectralGap}%
  \BibitemOpen
  \bibfield  {author} {\bibinfo {author} {\bibfnamefont {M.~B.}\ \bibnamefont
  {{Hastings}}}\ and\ \bibinfo {author} {\bibfnamefont {T.}~\bibnamefont
  {{Koma}}},\ }\bibfield  {title} {\enquote {\bibinfo {title} {{Spectral Gap
  and Exponential Decay of Correlations}},}\ }\href {\doibase
  10.1007/s00220-006-0030-4} {\bibfield  {journal} {\bibinfo  {journal}
  {Communications in Mathematical Physics}\ }\textbf {\bibinfo {volume}
  {265}},\ \bibinfo {pages} {781--804} (\bibinfo {year} {2006})},\ \Eprint
  {http://arxiv.org/abs/math-ph/0507008} {math-ph/0507008} \BibitemShut
  {NoStop}%
\bibitem [{\citenamefont {Roberts}\ and\ \citenamefont
  {Swingle}(2016)}]{RobertsSwingle}%
  \BibitemOpen
  \bibfield  {author} {\bibinfo {author} {\bibfnamefont {Daniel~A.}\
  \bibnamefont {Roberts}}\ and\ \bibinfo {author} {\bibfnamefont {Brian}\
  \bibnamefont {Swingle}},\ }\bibfield  {title} {\enquote {\bibinfo {title}
  {Lieb-robinson bound and the butterfly effect in quantum field theories},}\
  }\href {\doibase 10.1103/PhysRevLett.117.091602} {\bibfield  {journal}
  {\bibinfo  {journal} {Phys. Rev. Lett.}\ }\textbf {\bibinfo {volume} {117}},\
  \bibinfo {pages} {091602} (\bibinfo {year} {2016})}\BibitemShut {NoStop}%
\bibitem [{Note2()}]{Note2}%
  \BibitemOpen
  \bibinfo {note} {Here $\protect \mathaccentV {tilde}07E{v}_B({\protect \bf
  \protect \mathaccentV {hat}05En})$, with a tilde, denotes the normal
  propagation speed of a straight front whose normal is parallel to ${\protect
  \bf \protect \mathaccentV {hat}05En}$. In Ref.~\cite {opspreadAdam} this was
  denoted $v_B({\protect \bf \protect \mathaccentV {hat}05En})$, but here we
  use $v_B({\protect \bf \protect \mathaccentV {hat}05En})$ to denote the speed
  at which an initially local operator spreads away from the origin in the
  direction ${\protect \bf \protect \mathaccentV {hat}05En}$. These differ
  because in the absence of rotational symmetry the operator's front is not in
  general perpendicular to the radial vector, but they are related by a
  geometrical construction known from classical droplet growth \cite
  {WolfWulff, krug1991solids,opspreadAdam}.}\BibitemShut {Stop}%
\bibitem [{\citenamefont {{Das}}\ \emph {et~al.}(2017)\citenamefont {{Das}},
  \citenamefont {{Chakrabarty}}, \citenamefont {{Dhar}}, \citenamefont
  {{Kundu}}, \citenamefont {{Moessner}}, \citenamefont {{Sankar Ray}},\ and\
  \citenamefont {{Bhattacharjee}}}]{DharClassicalSpinChainChaos}%
  \BibitemOpen
  \bibfield  {author} {\bibinfo {author} {\bibfnamefont {A.}~\bibnamefont
  {{Das}}}, \bibinfo {author} {\bibfnamefont {S.}~\bibnamefont
  {{Chakrabarty}}}, \bibinfo {author} {\bibfnamefont {A.}~\bibnamefont
  {{Dhar}}}, \bibinfo {author} {\bibfnamefont {A.}~\bibnamefont {{Kundu}}},
  \bibinfo {author} {\bibfnamefont {R.}~\bibnamefont {{Moessner}}}, \bibinfo
  {author} {\bibfnamefont {S.}~\bibnamefont {{Sankar Ray}}}, \ and\ \bibinfo
  {author} {\bibfnamefont {S.}~\bibnamefont {{Bhattacharjee}}},\ }\bibfield
  {title} {\enquote {\bibinfo {title} {{Light-cone spreading of perturbations
  and the butterfly effect in a classical spin chain}},}\ }\href@noop {}
  {\bibfield  {journal} {\bibinfo  {journal} {ArXiv e-prints}\ } (\bibinfo
  {year} {2017})},\ \Eprint {http://arxiv.org/abs/1711.07505} {arXiv:1711.07505
  [cond-mat.stat-mech]} \BibitemShut {NoStop}%
\bibitem [{\citenamefont {Livi}\ \emph {et~al.}(1992)\citenamefont {Livi},
  \citenamefont {Politi},\ and\ \citenamefont {Ruffo}}]{livi1992scaling}%
  \BibitemOpen
  \bibfield  {author} {\bibinfo {author} {\bibfnamefont {R}~\bibnamefont
  {Livi}}, \bibinfo {author} {\bibfnamefont {A}~\bibnamefont {Politi}}, \ and\
  \bibinfo {author} {\bibfnamefont {S}~\bibnamefont {Ruffo}},\ }\bibfield
  {title} {\enquote {\bibinfo {title} {Scaling-law for the maximal lyapunov
  exponent},}\ }\href@noop {} {\bibfield  {journal} {\bibinfo  {journal}
  {Journal of Physics A: Mathematical and General}\ }\textbf {\bibinfo {volume}
  {25}},\ \bibinfo {pages} {4813} (\bibinfo {year} {1992})}\BibitemShut
  {NoStop}%
\bibitem [{\citenamefont {Kaneko}(1992)}]{kaneko1992propagation}%
  \BibitemOpen
  \bibfield  {author} {\bibinfo {author} {\bibfnamefont {Kunihiko}\
  \bibnamefont {Kaneko}},\ }\bibfield  {title} {\enquote {\bibinfo {title}
  {Propagation of disturbance, co-moving lyapunov exponent and path
  summation},}\ }\href@noop {} {\bibfield  {journal} {\bibinfo  {journal}
  {Physics Letters A}\ }\textbf {\bibinfo {volume} {170}},\ \bibinfo {pages}
  {210--216} (\bibinfo {year} {1992})}\BibitemShut {NoStop}%
\bibitem [{\citenamefont {Pikovsky}\ and\ \citenamefont
  {Kurths}(1994)}]{pikovsky1994roughening}%
  \BibitemOpen
  \bibfield  {author} {\bibinfo {author} {\bibfnamefont {Arkady~S}\
  \bibnamefont {Pikovsky}}\ and\ \bibinfo {author} {\bibfnamefont {J{\"u}rgen}\
  \bibnamefont {Kurths}},\ }\bibfield  {title} {\enquote {\bibinfo {title}
  {Roughening interfaces in the dynamics of perturbations of spatiotemporal
  chaos},}\ }\href@noop {} {\bibfield  {journal} {\bibinfo  {journal} {Physical
  Review E}\ }\textbf {\bibinfo {volume} {49}},\ \bibinfo {pages} {898}
  (\bibinfo {year} {1994})}\BibitemShut {NoStop}%
\bibitem [{Note3()}]{Note3}%
  \BibitemOpen
  \bibinfo {note} {The expression on the left-hand side of (\ref
  {eq:classicalquantity}) becomes a ``partition function'' for two paths. The
  local weights $\partial u({\protect \bf y}_{i+1}, i+1) / \partial u({\protect
  \bf y}_{i}, i)$ depend not only on ${\protect \bf y}_{i+1}$ and ${\protect
  \bf y}_{i}$ but also on the configuration $u({\protect \bf y_i},i)$. The
  chaotic time-dependence of $u({\protect \bf y}_i,i)$ means that the
  configurational average has a similar effect to averaging over weakly
  correlated randomness in the weights. Since we are averaging the ``partition
  function'', rather than its logarithm, this is an annealed average, and
  $-\lambda ({\protect \bf v})t$ is an annealed ``free energy'' for the pair of
  paths. The quenched free energy, in which we take the logarithm before
  averaging, would give the more conventional definition of the Lyapunov
  exponent \cite
  {livi1992scaling,kaneko1992propagation,pikovsky1994roughening}.}\BibitemShut
  {Stop}%
\bibitem [{\citenamefont {Nahum}\ \emph {et~al.}(2017)\citenamefont {Nahum},
  \citenamefont {Ruhman}, \citenamefont {Vijay},\ and\ \citenamefont
  {Haah}}]{nahum2017quantum}%
  \BibitemOpen
  \bibfield  {author} {\bibinfo {author} {\bibfnamefont {Adam}\ \bibnamefont
  {Nahum}}, \bibinfo {author} {\bibfnamefont {Jonathan}\ \bibnamefont
  {Ruhman}}, \bibinfo {author} {\bibfnamefont {Sagar}\ \bibnamefont {Vijay}}, \
  and\ \bibinfo {author} {\bibfnamefont {Jeongwan}\ \bibnamefont {Haah}},\
  }\bibfield  {title} {\enquote {\bibinfo {title} {Quantum entanglement growth
  under random unitary dynamics},}\ }\href@noop {} {\bibfield  {journal}
  {\bibinfo  {journal} {Physical Review X}\ }\textbf {\bibinfo {volume} {7}},\
  \bibinfo {pages} {031016} (\bibinfo {year} {2017})}\BibitemShut {NoStop}%
\bibitem [{Note4()}]{Note4}%
  \BibitemOpen
  \bibinfo {note} {{Inside the light cone there is a large deviation form
  governing convergence to the saturation value: ${C_{1d}^{\protect \rm
  rc}(x,t) \sim 1- \protect \qopname \relax o{exp}{\setbox \z@ \hbox
  {\frozen@everymath \@emptytoks \mathsurround \z@ $\nulldelimiterspace \z@
  \left (\vcenter to\@ne \big@size {}\right .$}\box \z@ }{-\protect \frac
  {(v-v_B)^2}{2D} t}{\setbox \z@ \hbox {\frozen@everymath \@emptytoks
  \mathsurround \z@ $\nulldelimiterspace \z@ \left )\vcenter to\@ne \big@size
  {}\right .$}\box \z@ }}$. The exponent here is the continuation of $\lambda
  (v)$ outside the front. However, in the higher dimensional examples, the
  large deviation form inside the front scales with a distinct power of $t$,
  $t^d$ in $d$ spatial dimensions \cite {MajumdarKPZTail}. In the presence of
  additional conserved densities (like energy or charge), the late time
  saturation of the OTOC is a power-law in time instead of exponential~\cite
  {KhemaniCons, TiborCons}.}}\BibitemShut {Stop}%
\bibitem [{Note5()}]{Note5}%
  \BibitemOpen
  \bibinfo {note} {{ In random circuits related random walk pictures underlie
  the calculation of both the OTOC and the second Renyi entropy \cite
  {opspreadAdam, opspreadCurt}. In these random systems this yields a relation
  between $\lambda (v)$ and the ``entanglement line tension'' defined in \cite
  {jonay}, specifically the line tension $\protect \mathcal {E}_2(v)$ for the
  second Renyi entropy. This motivates the conjecture, for non-random systems,
  that $\lambda (v)|_\protect \text {cont} = - s_\protect \text {eq} ( \protect
  \mathcal {E}_2(v) - v)$, where $s_\protect \text {eq}$ is the thermal entropy
  density. The left hand side denotes the analytic continuation of $\lambda
  (v)$ from ${v>v_B}$ to values $v<v_B$. In random circuits we must distinguish
  different kinds of averages. The line tension extracted from a calculation of
  $\protect \overline {e^{-S_2}}$ determines $\lambda (v)$ for the average OTOC
  $\protect \overline {C(x,t)}$ by the above formula. It is natural to expect
  that the line tension determined by the more natural direct average $\protect
  \overline {S_2}$ determines $\lambda (v)$ for the typical value of the OTOC,
  $\protect \qopname \relax o{exp}\protect \overline {\protect \qopname \relax
  o{ln}C(x,t)}$. The average and typical values of the OTOC are parametrically
  close in the region close to the front, but they may differ significantly in
  the far-front regime where both are exponentially small.}}\BibitemShut
  {Stop}%
\bibitem [{Note6()}]{Note6}%
  \BibitemOpen
  \bibinfo {note} {{In some circuit models in $d>1$ (which do not have
  continuous spatial rotation symmetry) some sections of the operator's front
  can be ``glued'' to the strict lightcone defined by the discrete time circuit
  \cite {opspreadAdam}. This is a peculiar case where ${v_B( {\protect \bf
  \protect \mathaccentV {hat}05En} ) = v_\protect \text {LC}( {\protect \bf
  \protect \mathaccentV {hat}05En})}$ for some directions ${\protect \bf
  \protect \mathaccentV {hat}05En}$ in space, so that no nontrivial $\lambda
  ({\protect \mathbf v})$ can be defined for these directions of ${\protect \bf
  v}$.}}\BibitemShut {Stop}%
\bibitem [{\citenamefont {{Le Doussal}}\ \emph {et~al.}(2016)\citenamefont {{Le
  Doussal}}, \citenamefont {{Majumdar}},\ and\ \citenamefont
  {{Schehr}}}]{MajumdarKPZTail}%
  \BibitemOpen
  \bibfield  {author} {\bibinfo {author} {\bibfnamefont {P.}~\bibnamefont {{Le
  Doussal}}}, \bibinfo {author} {\bibfnamefont {S.~N.}\ \bibnamefont
  {{Majumdar}}}, \ and\ \bibinfo {author} {\bibfnamefont {G.}~\bibnamefont
  {{Schehr}}},\ }\bibfield  {title} {\enquote {\bibinfo {title} {{Large
  deviations for the height in 1D Kardar-Parisi-Zhang growth at late times}},}\
  }\href {\doibase 10.1209/0295-5075/113/60004} {\bibfield  {journal} {\bibinfo
   {journal} {EPL (Europhysics Letters)}\ }\textbf {\bibinfo {volume} {113}},\
  \bibinfo {pages} {60004} (\bibinfo {year} {2016})},\ \Eprint
  {http://arxiv.org/abs/1601.05957} {arXiv:1601.05957 [cond-mat.stat-mech]}
  \BibitemShut {NoStop}%
\bibitem [{\citenamefont {{Monthus}}\ and\ \citenamefont
  {{Garel}}(2006)}]{MonthusKPZNumerics}%
  \BibitemOpen
  \bibfield  {author} {\bibinfo {author} {\bibfnamefont {C.}~\bibnamefont
  {{Monthus}}}\ and\ \bibinfo {author} {\bibfnamefont {T.}~\bibnamefont
  {{Garel}}},\ }\bibfield  {title} {\enquote {\bibinfo {title} {{Probing the
  tails of the ground-state energy distribution for the directed polymer in a
  random medium of dimension d=1,2,3 via a Monte Carlo procedure in the
  disorder}},}\ }\href {\doibase 10.1103/PhysRevE.74.051109} {\bibfield
  {journal} {\bibinfo  {journal} {\pre}\ }\textbf {\bibinfo {volume} {74}},\
  \bibinfo {eid} {051109} (\bibinfo {year} {2006})},\ \Eprint
  {http://arxiv.org/abs/cond-mat/0607411} {cond-mat/0607411} \BibitemShut
  {NoStop}%
\bibitem [{\citenamefont {{Kolokolov}}\ and\ \citenamefont
  {{Korshunov}}(2008)}]{KolokolovKPZtails}%
  \BibitemOpen
  \bibfield  {author} {\bibinfo {author} {\bibfnamefont {I.~V.}\ \bibnamefont
  {{Kolokolov}}}\ and\ \bibinfo {author} {\bibfnamefont {S.~E.}\ \bibnamefont
  {{Korshunov}}},\ }\bibfield  {title} {\enquote {\bibinfo {title} {{Universal
  and nonuniversal tails of distribution functions in the directed polymer and
  Kardar-Parisi-Zhang problems}},}\ }\href {\doibase
  10.1103/PhysRevB.78.024206} {\bibfield  {journal} {\bibinfo  {journal}
  {\prb}\ }\textbf {\bibinfo {volume} {78}},\ \bibinfo {eid} {024206} (\bibinfo
  {year} {2008})},\ \Eprint {http://arxiv.org/abs/0805.0402} {arXiv:0805.0402
  [cond-mat.dis-nn]} \BibitemShut {NoStop}%
\bibitem [{\citenamefont {Pagnani}\ and\ \citenamefont
  {Parisi}(2015)}]{KPZ_2d_exponent}%
  \BibitemOpen
  \bibfield  {author} {\bibinfo {author} {\bibfnamefont {Andrea}\ \bibnamefont
  {Pagnani}}\ and\ \bibinfo {author} {\bibfnamefont {Giorgio}\ \bibnamefont
  {Parisi}},\ }\bibfield  {title} {\enquote {\bibinfo {title} {Numerical
  estimate of the kardar-parisi-zhang universality class in (2+1)
  dimensions},}\ }\href {\doibase 10.1103/PhysRevE.92.010101} {\bibfield
  {journal} {\bibinfo  {journal} {Phys. Rev. E}\ }\textbf {\bibinfo {volume}
  {92}},\ \bibinfo {pages} {010101} (\bibinfo {year} {2015})}\BibitemShut
  {NoStop}%
\bibitem [{\citenamefont {Kelling}\ \emph {et~al.}(2016)\citenamefont
  {Kelling}, \citenamefont {\'Odor},\ and\ \citenamefont
  {Gemming}}]{KellingKPZ_2d}%
  \BibitemOpen
  \bibfield  {author} {\bibinfo {author} {\bibfnamefont {Jeffrey}\ \bibnamefont
  {Kelling}}, \bibinfo {author} {\bibfnamefont {G\'eza}\ \bibnamefont
  {\'Odor}}, \ and\ \bibinfo {author} {\bibfnamefont {Sibylle}\ \bibnamefont
  {Gemming}},\ }\bibfield  {title} {\enquote {\bibinfo {title} {Universality of
  (2+1)-dimensional restricted solid-on-solid models},}\ }\href {\doibase
  10.1103/PhysRevE.94.022107} {\bibfield  {journal} {\bibinfo  {journal} {Phys.
  Rev. E}\ }\textbf {\bibinfo {volume} {94}},\ \bibinfo {pages} {022107}
  (\bibinfo {year} {2016})}\BibitemShut {NoStop}%
\bibitem [{\citenamefont {Halpin-Healy}(2012)}]{HealyScaling1}%
  \BibitemOpen
  \bibfield  {author} {\bibinfo {author} {\bibfnamefont {Timothy}\ \bibnamefont
  {Halpin-Healy}},\ }\bibfield  {title} {\enquote {\bibinfo {title}
  {($2\mathbf{+}1$)-dimensional directed polymer in a random medium: Scaling
  phenomena and universal distributions},}\ }\href {\doibase
  10.1103/PhysRevLett.109.170602} {\bibfield  {journal} {\bibinfo  {journal}
  {Phys. Rev. Lett.}\ }\textbf {\bibinfo {volume} {109}},\ \bibinfo {pages}
  {170602} (\bibinfo {year} {2012})}\BibitemShut {NoStop}%
\bibitem [{\citenamefont {Halpin-Healy}(2013)}]{HealyScaling2}%
  \BibitemOpen
  \bibfield  {author} {\bibinfo {author} {\bibfnamefont {Timothy}\ \bibnamefont
  {Halpin-Healy}},\ }\bibfield  {title} {\enquote {\bibinfo {title} {Extremal
  paths, the stochastic heat equation, and the three-dimensional
  kardar-parisi-zhang universality class},}\ }\href {\doibase
  10.1103/PhysRevE.88.042118} {\bibfield  {journal} {\bibinfo  {journal} {Phys.
  Rev. E}\ }\textbf {\bibinfo {volume} {88}},\ \bibinfo {pages} {042118}
  (\bibinfo {year} {2013})}\BibitemShut {NoStop}%
\bibitem [{\citenamefont {{Halpin-Healy}}\ and\ \citenamefont
  {{Takeuchi}}(2015)}]{KPZcocktail}%
  \BibitemOpen
  \bibfield  {author} {\bibinfo {author} {\bibfnamefont {T.}~\bibnamefont
  {{Halpin-Healy}}}\ and\ \bibinfo {author} {\bibfnamefont {K.~A.}\
  \bibnamefont {{Takeuchi}}},\ }\bibfield  {title} {\enquote {\bibinfo {title}
  {{A KPZ Cocktail-Shaken, not Stirred...}}}\ }\href {\doibase
  10.1007/s10955-015-1282-1} {\bibfield  {journal} {\bibinfo  {journal}
  {Journal of Statistical Physics}\ }\textbf {\bibinfo {volume} {160}},\
  \bibinfo {pages} {794--814} (\bibinfo {year} {2015})},\ \Eprint
  {http://arxiv.org/abs/1505.01910} {arXiv:1505.01910 [cond-mat.stat-mech]}
  \BibitemShut {NoStop}%
\bibitem [{Jac()}]{JacobLesik_private}%
  \BibitemOpen
  \href@noop {} {}\bibinfo {note} {Cheng-Ju Lin, Olexei Motrunich, private
  communication.}\BibitemShut {Stop}%
\bibitem [{Dha()}]{Dhar_unpublished}%
  \BibitemOpen
  \href@noop {} {}\bibinfo {note} {Abhishek Dhar, unpublished}\BibitemShut
  {NoStop}%
\bibitem [{\citenamefont {{Gopalakrishnan}}\ \emph {et~al.}(2018)\citenamefont
  {{Gopalakrishnan}}, \citenamefont {{Huse}}, \citenamefont {{Khemani}},\ and\
  \citenamefont {{Vasseur}}}]{ghkv}%
  \BibitemOpen
  \bibfield  {author} {\bibinfo {author} {\bibfnamefont {S.}~\bibnamefont
  {{Gopalakrishnan}}}, \bibinfo {author} {\bibfnamefont {D.~A.}\ \bibnamefont
  {{Huse}}}, \bibinfo {author} {\bibfnamefont {V.}~\bibnamefont {{Khemani}}}, \
  and\ \bibinfo {author} {\bibfnamefont {R.}~\bibnamefont {{Vasseur}}},\
  }\bibfield  {title} {\enquote {\bibinfo {title} {{Hydrodynamics of operator
  spreading and quasiparticle diffusion in interacting integrable systems}},}\
  }\href@noop {} {\bibfield  {journal} {\bibinfo  {journal} {ArXiv e-prints}\ }
  (\bibinfo {year} {2018})},\ \Eprint {http://arxiv.org/abs/1809.02126}
  {arXiv:1809.02126 [cond-mat.stat-mech]} \BibitemShut {NoStop}%
\bibitem [{Note7()}]{Note7}%
  \BibitemOpen
  \bibinfo {note} {Let the probability distribution for weak-link ``waiting
  times'' be ${P(\tau )\sim \tau ^{-a-2}}$. At weak disorder $(1<a)$ the
  broadening of the operator's front \cite {NahumRuhmanHuse} is diffusive, as
  in the clean system. At intermediate disorder (${0<a<1}$) the front broadens
  more strongly, giving ${\lambda (v)\sim -(v-v_B)^{(a+1)/a}}$. For strong
  disorder ($-1<a<0$) the butterfly speed vanishes: in this regime ${\lambda
  (v)\sim - |v|^{1-|a|}}$. In the disordered system the definition of $\lambda
  (v)$ depends on whether we consider e.g. the mean or the typical value of the
  OTOC, but this should not change these exponents.}\BibitemShut {Stop}%
\bibitem [{\citenamefont {{Xu}}\ and\ \citenamefont
  {{Swingle}}(2018)}]{Swingle_otocMPS}%
  \BibitemOpen
  \bibfield  {author} {\bibinfo {author} {\bibfnamefont {S.}~\bibnamefont
  {{Xu}}}\ and\ \bibinfo {author} {\bibfnamefont {B.}~\bibnamefont
  {{Swingle}}},\ }\bibfield  {title} {\enquote {\bibinfo {title} {{Accessing
  scrambling using matrix product operators}},}\ }\href@noop {} {\bibfield
  {journal} {\bibinfo  {journal} {ArXiv e-prints}\ } (\bibinfo {year}
  {2018})},\ \Eprint {http://arxiv.org/abs/1802.00801} {arXiv:1802.00801
  [quant-ph]} \BibitemShut {NoStop}%
\bibitem [{\citenamefont {Wolf}(1987)}]{WolfWulff}%
  \BibitemOpen
  \bibfield  {author} {\bibinfo {author} {\bibfnamefont {D~E}\ \bibnamefont
  {Wolf}},\ }\bibfield  {title} {\enquote {\bibinfo {title} {Wulff construction
  and anisotropic surface properties of two-dimensional eden clusters},}\
  }\href {http://stacks.iop.org/0305-4470/20/i=5/a=033} {\bibfield  {journal}
  {\bibinfo  {journal} {Journal of Physics A: Mathematical and General}\
  }\textbf {\bibinfo {volume} {20}},\ \bibinfo {pages} {1251} (\bibinfo {year}
  {1987})}\BibitemShut {NoStop}%
\bibitem [{\citenamefont {Krug}\ \emph {et~al.}(1991)\citenamefont {Krug},
  \citenamefont {Spohn},\ and\ \citenamefont {Godr{\`e}che}}]{krug1991solids}%
  \BibitemOpen
  \bibfield  {author} {\bibinfo {author} {\bibfnamefont {J}~\bibnamefont
  {Krug}}, \bibinfo {author} {\bibfnamefont {H}~\bibnamefont {Spohn}}, \ and\
  \bibinfo {author} {\bibfnamefont {C}~\bibnamefont {Godr{\`e}che}},\
  }\bibfield  {title} {\enquote {\bibinfo {title} {in `solids far from
  equilibrium'},}\ }\href@noop {} {\bibfield  {journal} {\bibinfo  {journal}
  {Solids far from equilibrium}\ } (\bibinfo {year} {1991})}\BibitemShut
  {NoStop}%
\end{thebibliography}%

\begin{appendix}

\end{appendix}

\end{document}